\documentclass[twocolumn,final,english,prl,notitlepage,nofootinbib,floatfix,longbibliography]{revtex4-1}

\usepackage[T1]{fontenc}
\usepackage[utf8]{inputenc}
\usepackage{refstyle}
\usepackage{amsmath}
\usepackage{bm} 
\usepackage{amssymb}
\usepackage{graphicx}
\usepackage[pdftex, hidelinks]{hyperref}
\usepackage[usenames,dvipsnames]{xcolor}
\usepackage{bbm}
\usepackage{tikz}
\usepackage{xspace}
\usepackage{bm}
\usepackage{booktabs}
\usepackage{babel}
\usepackage[separate-uncertainty=true]{siunitx}
\usepackage[makeroom]{cancel}

\usepackage{ifdraft}

\usepackage[normalem]{ulem} 

\usetikzlibrary{matrix}

\newcommand\myshade{85}
\colorlet{mylinkcolor}{violet}
\colorlet{mycitecolor}{YellowOrange}
\colorlet{myurlcolor}{Aquamarine}
\hypersetup{
  linkcolor  = mylinkcolor!\myshade!black,
  citecolor  = mycitecolor!\myshade!black,
  urlcolor   = myurlcolor!\myshade!black,
  colorlinks = true,
}

\DeclareMathAlphabet\mathbfcal{OMS}{cmsy}{b}{n}

\makeatletter

\def\@fnsymbol#1{\ensuremath{\ifcase#1\or *\or *,\dagger\or \ddagger\or
   \mathsection\or \mathparagraph\or \|\or *\or \dagger\dagger
   \or \ddagger\ddagger \else\@ctrerr\fi}}

\AtBeginDocument{}
\RS@ifundefined{subsecref}
  {\newref{subsec}{name = \RSsectxt}}
  {}
\RS@ifundefined{thmref}
  {\def\RSthmtxt{theorem~}\newref{thm}{name = \RSthmtxt}}
  {}
\RS@ifundefined{lemref}
  {\def\RSlemtxt{lemma~}\newref{lem}{name = \RSlemtxt}}
  {}

\makeatother

\newcommand{\sref}[2]{\hyperref[#1]{\ref*{#1}(#2)}}

\newcommand{\epr}{\ensuremath{\mathrm{\scriptscriptstyle EPR}}}

\newcommand{\LO}[1]{\ensuremath{\mathrm{LO_{#1}}}}

\newcommand{\Vc}{\ensuremath{V_\mathrm{c}}\xspace}
\newcommand{\Vu}{\ensuremath{V_\mathrm{u}}\xspace}

\newcommand{\Tr}{\mathrm{Tr}}

\newcommand{\Xeprc}{X_\epr^c}
\newcommand{\Peprc}{P_\epr^c}
\newcommand{\Xeprcrot}{\tilde{X}_\epr^c}
\newcommand{\Peprcrot}{\tilde{P}_\epr^c}

\newcommand{\Cq}{\ensuremath{C_\mathrm{q}}\xspace}

\DeclareMathOperator{\Var}{Var}

\newcommand\resonantVc{0.83 \pm 0.02}
\newcommand\resonantVu{1.91 \pm 0.05}
\newcommand\detunedVc{2.02 \pm 0.03}

\newcommand\resonantVune{1.91}

\newcommand\detunedVune{6.07}
\newcommand\resonantBAreduction{\SI{4.6}{\decibel}\xspace}
\newcommand\GammaS{\SI{20}{\kHz}}
\newcommand\gammaS{\SI{2.9}{\kHz}}
\newcommand\gammaSzero{\SI{1.7}{\kHz}}

\newcommand\nS{0.8}
\newcommand\avgDelta{\SI{-0.7}{\MHz}}
\newcommand\GammaM{\SI{15}{\kHz}}
\newcommand\gammaM{\SI{3.9}{\kHz}}
\newcommand\spring{\SI{1}{\kHz}}
\newcommand\effnu{0.53}
\newcommand\effeta{0.77}
\newcommand\ovc{93}
\newcommand\CqM{15}
\newcommand\QBATHM{19}
\newcommand\QBATHS{4.9}
\newcommand{\atomicPol}{\ensuremath{{0.82 \pm 0.01}}}
\newcommand\jointThermalReduction{\ensuremath{\SI{2.5}{\decibel}}}
\newcommand\BBNoiseLevel{1.68}
\newcommand\gammaSratio{0.6}
\newcommand\Omegamzero{1.370}
\newcommand\Omegam{1.369}\xspace
\newcommand\mechT{11}
\newcommand\kappamech{4.2}
\newcommand\gammamechnat{2.1}
\newcommand\resonantVureduction{5.0}
\newcommand\nthmech{173}
\newcommand\zetaS{0.028}

\newcommand\nM{\ensuremath{\sim2}}

\begin{document}

\title{Entanglement between Distant Macroscopic Mechanical and Spin Systems}

\author{Rodrigo A. Thomas}
\thanks{These authors contributed equally to the work}
\affiliation{Niels Bohr Institute, University of Copenhagen, Copenhagen, Denmark}
\author{Michał Parniak}
\thanks{These authors contributed equally to the work}
\affiliation{Niels Bohr Institute, University of Copenhagen, Copenhagen, Denmark}
\author{Christoffer Østfeldt}
\thanks{These authors contributed equally to the work}
\affiliation{Niels Bohr Institute, University of Copenhagen, Copenhagen, Denmark}
\author{Chistoffer B. Møller}
\thanks{These authors contributed equally to the work}
\altaffiliation[Currently at: ]{ICFO -- Institut De Ciencies Fotoniques, The Barcelona Institute of Science and Technology, Castelldefels, Spain}
\affiliation{Niels Bohr Institute, University of Copenhagen, Copenhagen, Denmark}
\author{Christian Bærentsen}
\affiliation{Niels Bohr Institute, University of Copenhagen, Copenhagen, Denmark}
\author{Yeghishe Tsaturyan}
\altaffiliation[Currently at: ]{Pritzker School of Molecular Engineering, University of Chicago, Chicago, USA }
\affiliation{Niels Bohr Institute, University of Copenhagen, Copenhagen, Denmark}
\author{Albert Schliesser}
\affiliation{Niels Bohr Institute, University of Copenhagen, Copenhagen, Denmark}
\author{Jürgen Appel}
\altaffiliation[Currently at: ]{Danish Fundamental Metrology, Hørsholm, Denmark}
\affiliation{Niels Bohr Institute, University of Copenhagen, Copenhagen, Denmark}
\author{Emil Zeuthen}
\affiliation{Niels Bohr Institute, University of Copenhagen, Copenhagen, Denmark}
\author{Eugene S. Polzik}
\altaffiliation[Corresponding author: ]{polzik@nbi.ku.dk}
\affiliation{Niels Bohr Institute, University of Copenhagen, Copenhagen, Denmark}

\begin{abstract}
Entanglement is a vital property of multipartite quantum systems, characterised by the inseparability of quantum states of objects regardless of their spatial separation. Generation of entanglement between increasingly macroscopic and disparate systems is an ongoing effort in quantum science which enables hybrid quantum networks~\cite{KimbleNature2008,Kurizki2015}, quantum-enhanced sensing~\cite{Degen2017}, and probing the fundamental limits of quantum theory~\citep{Chen2013,cavityoptomechRMP}. The disparity of hybrid systems and the vulnerability of quantum correlations have thus far hampered the generation of macroscopic hybrid entanglement. Here we demonstrate, for the first time, generation of an entangled state between the motion of a macroscopic mechanical oscillator and a collective atomic spin oscillator, as witnessed by an Einstein-Podolsky-Rosen variance below the separability limit~\cite{Duan2000}, $\resonantVc < 1$. The mechanical oscillator is a millimeter-size dielectric membrane~\cite{softclampingTsaturyan2017} and the spin oscillator is an ensemble of $10^9$ atoms in a magnetic field~\cite{Borregaard2016}. Light propagating through the two spatially separated systems generates entanglement due to the collective spin playing the role of an effective negative-mass reference frame~\citep{julsgaard2001experimental, eprpolzikhammerer2009, polzikadp2015,Moller2017} and providing, under ideal circumstances, a backaction-free subspace~\citep{tsangcavesprl2010}; in the experiment, quantum backaction is suppressed by $\resonantBAreduction$. Our results pave the road towards measurement of motion beyond the standard quantum limits of sensitivity with applications in force, acceleration, and gravitational wave detection~\cite{KhaliliESPprl2018,Zeuthenpolzikhaliliprd2019}, as well as towards teleportation-based protocols~\cite{RMPentanglement2009} in hybrid quantum networks.
\end{abstract}

\maketitle

Entanglement is a key resource for quantum information processing. 
In particular, entangled states of motional and spin degrees of freedom have played a prominent role in quantum computing and simulation with trapped ions and atoms~\citep{CiracZoller95,GrossBlochScience2017,BrownKimMonroenpjQI2016}.
There, entanglement between motion and spin is generated by short-range interactions between individual atoms positioned at micron-scale distances, with motional and spin degrees of freedom associated with the same atoms. 

A very different regime, focused on long-range macroscopic entanglement between the motion of one object and a spin of another, has been proposed in Ref.~\citep{eprpolzikhammerer2009}. The key idea is that an atomic spin in a magnetic field acts as a negative-mass oscillator, permitting travelling light to generate entanglement between the two objects. The negative-mass idea, which has been implicitly used in earlier experiments
with two atomic ensembles~\cite{julsgaard2001experimental,RevModPhys.82.1041,Muschik2011,Krauter2011}, has been further developed in Refs.~\citep{Stannigel2012,Vasilyev2013,huangzeuthenprl2018,Manukhova2020} and has become the basis for quantum-mechanics-free subspaces~\citep{cavestsangprx2012}. 

Negative-mass-enabled instability~\citep{stamperkurn-2018} and quantum backaction (QBA) evasion~\cite{Moller2017} have been recently demonstrated using the coupling of a motional degree of freedom to a spin system. In Refs.~\citep{jockel2015sympathetic,christoph2018combined} sympathetic cooling of a mechanical oscillator optically coupled to atoms have been shown. The negative-mass reference frame idea has also been utilized in proposals~\cite{Tan2013,WoolleyClerkPRA2013} by using an auxiliary mechanical system and multiple drive tones. In this way, entanglement has been generated between two micromechanical oscillators embedded in a common microwave cavity~\citep{Ockeloen-KorppiNature2018}. An approach to mechanical-mechanical entanglement based on single-photon detection was demonstrated in Refs.~\cite{RiedingerNature2018,LeeScience2011}.

\begin{figure}[tbp]
    \centering
    \includegraphics[width=\columnwidth]{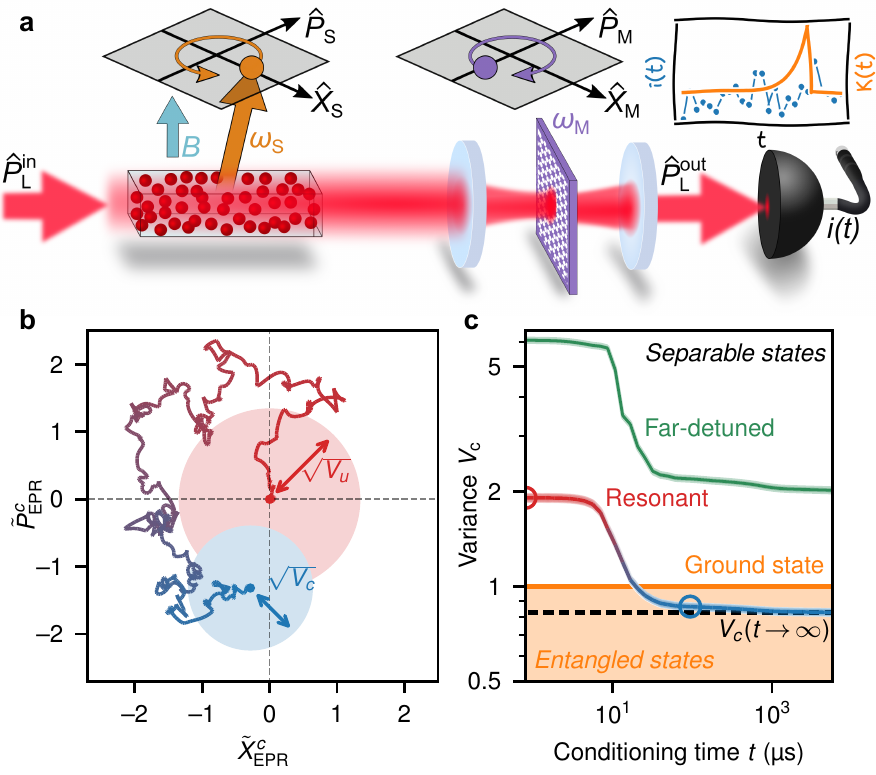}
    \caption{\textbf{Tracking of the Einstein-Podolsky-Rosen oscillator}. \textbf{a,} A simplified schematic of the entangled system, consisting of an atomic spin ensemble and a mechanical oscillator in a cavity, separated from the atoms by a one meter distance, and probed by light in a cascaded manner. The measurement photocurrent $i(t)$ is convolved with a Wiener filter $K(t',t)$ (approximate envelope shown in the inset), to yield a conditional trajectory. \textbf{b,} Quantum  phase-space trajectory of an EPR-entangled oscillator pair along with deterministic variance of the estimate $V_u=\resonantVu$ for $t=0$ (red), and the approximately final conditional variance of $V_c=\resonantVc$ at $t=\SI{110}{\micro\second}$ (blue). 
    \textbf{c,} Evolution of the conditional variance $V_c$ for the resonant (red to blue) and far-detuned (green), i.e., for a joint and spectrally separated oscillators, respectively. The circle marks the variance at the end of the trajectory in \textbf{b}. The shaded areas mark the $1\sigma$ uncertainty of $V_c$.
    }
    \label{fig:trajectory}
\end{figure}

Here we report an experimental implementation of Einstein-Podolsky-Rosen~(EPR) entanglement in a hybrid system consisting of a mechanical oscillator and a spin oscillator~\cite{eprpolzikhammerer2009}, as depicted schematically in Fig.~\ref{fig:trajectory}a.
An out-of-plane vibrational mode of a soft-clamped, highly stressed dielectric membrane~\citep{softclampingTsaturyan2017}, which is embedded in a free-space optical cavity, constitues the mechanical subsystem.
The spin subsystem is prepared in a warm ensemble of optically pumped caesium atoms confined in a spin-preserving microcell~\citep{Borregaard2016}. The two oscillators are coupled to an itinerant light field and optically read out in a cascaded fashion.

The collective macroscopic spin $\hat{J}_x=\sum_{i=1}^{N}\hat{F}_{x}^{(i)}$ of $N\approx10^9$ atoms, each with total angular momentum components $(\hat{F}_{x}^{(i)},\hat{F}_{y}^{(i)},\hat{F}_{z}^{(i)})$, is optically pumped in the direction $x$ of the magnetic bias field $B$.
In the limit where the magnitude of the mean longitudinal spin $J_x= |\langle \hat{J}_x \rangle|$ far exceeds the transverse collective spin components, $\hat{J}_y$ and $\hat{J}_z$, 
the latter can be mapped to the harmonic oscillator variables $\hat{X}_\text{S}=\hat{J}_z/\sqrt{\hbar J_x}$ and $\hat{P}_\text{S}=-\hat{J}_y/\sqrt{\hbar J_x}$, satisfying the canonical commutation relation $[\hat{X}_\text{S}, \hat{P}_\text{S}]=i$~\cite{holsteinprimakoff}. 
The transverse components precess around the magnetic field at the Larmor frequency $\omega_\text{S}\propto B$ according to $\hat{H}_\text{S}=-\hbar\omega_\text{S} \hat{J}_x \approx -\hbar\omega_\text{S} J_x + (\hbar \omega_\text{S}/2)(\hat{X}_{\text{S}}^{2}+\hat{P}_{\text{S}}^{2})$, where the first term is a constant offset. 

Since the optical pumping prepares the collective spin near the energetically highest Zeeman state, the collective spin realises a negative-mass oscillator, i.e., $\omega_{\textrm{S}} <0$~\citep{polzikadp2015}, with a counter-rotating trajectory~(see Fig.~\ref{fig:trajectory}a). The ``negative mass'' terminology arises by analogy to the standard harmonic oscillator Hamiltonian $\hat{H}=m\omega^2 \hat{X}^{2}/2 + \hat{P}^2/(2m)$, in which the sign of the mass $m$ determines that of both the potential and kinetic energies, as does the sign of $\omega_{\textrm{S}}$ in $\hat{H}_\text{S}$. 

Fundamentally, the non-commuting quadratures of motion $[\hat{X}_{j}(t),\hat{P}_{j}(t)]=i$ for the individual systems (where $j\in\{\text{S,M}\}$ labels spin and mechanics) cannot be known simultaneously with arbitrary precision due to the Heisenberg uncertainty principle; in particular, $\Var[\hat{X}_j]+\Var[\hat{P}_j]\geq 1$. This limit is enforced by the QBA of the meter field (e.g., light) on the measured oscillator.

Such a limit does not apply to a commuting combination of variables such as $[\hat{X}_{\epr},\hat{P}_{\epr}] \equiv [(\hat{X}_\text{M}-\hat{X}_\text{S})/\sqrt{2},(\hat{P}_\text{M}+\hat{P}_\text{S})/\sqrt{2}]=0$, i.e., the sum of variances is no longer bounded from below. 
In fact, 
$V=\Var[\hat{X}_{\epr}]+\Var[\hat{P}_{\epr}] < 1$~\cite{Duan2000} 
implies entanglement between systems $\text{S}$ and $\text{M}$, which is analogous to violating the single-system limit with the EPR variables. 
Since the EPR variables describe spatially separated systems, this effective oscillator is non-local.

We entangle the two oscillators by a backaction-evading collective position measurement. 
For matched frequencies, $-\omega_\text{S} = \omega_\text{M} \equiv \omega > 0$, the negative-mass spin oscillator's response to the perturbing optical field happens with a phase opposite to that of the positive-mass oscillator. The resulting information written onto the optical meter phase is
$\hat{P}_\text{L}^\text{out} \propto \hat{X}_{\epr}(t) = \cos\omega t \hat{X}_{\epr}(0) + \sin\omega t \hat{P}_{\epr}(0)$,
and thus only depends on the initial values of $\hat{X}_j$ and $\hat{P}_j$, in the absence of damping and intrinsic oscillator noise. Thus, under ideal conditions, the joint measurement on an EPR-entangled system produces a noiseless trajectory of one oscillator in the reference frame of the other~\cite{polzikadp2015}. 

In quantum theory, those trajectories arise as the expectation values of the dynamical variables with respect to the \emph{conditional} quantum state $\hat{\rho}_c(t)$, i.e., incorporating the information contained in the measurement record obtained at times $t'<t$. 
Tracking the conditional state evolution is relatively straightforward in the present case of Gaussian states, dynamics, and measurements (see Methods~\ref{app:wiener}), where $\hat{\rho}_c(t)$ is characterized solely by its first and second moments, which may be extracted by linear filtering of past measurement outcomes~\cite{YanbeiChenPRA2009,RossiPRL2019}. 
Optimal filter functions are determined from  the equations of motion, noise statistics, and the input-output relations for the light fields. 
The optimal filter that takes into account data from a time period $[0,t]$ to estimate, e.g., $\hat{X}_{\epr}$ is called the Wiener filter, 
a version of the Kalman filter widely used for state estimation~\citep{WieczorekPRL2015}. In the simplest case, the filter envelope is an exponential with the rate defined by decoherence and readout processes, as pictorially shown in the inset in Fig.~\ref{fig:trajectory}a. Such exponential filtering has been used in, e.g., Refs.~\cite{Wasilewski2010,Krauter2011}. From the Wiener filter $K_{X}$ for $\hat{X}_{\epr}$ the conditional quadrature is obtained as
\begin{equation}\label{eq:x-est}
\Xeprc(t) = \int_{0}^{t} \mathrm{d}t'\; K_{X} (t'-t,t) i(t'),
\end{equation}
where $i(t)$ is the instantaneous photocurrent obtained by the homodyne detection of the optical quadrature $\hat{P}_{\text{L}}^{\text{out}}(t)$ of the transmitted light. To obtain the exact Wiener filter, we solve the Wiener-Hopf equations (see Methods~\ref{app:wiener}), which involve the cross-correlation $C_{Xi}(t)$  between the oscillator signal $\hat{X}_{\epr}$ and $i(t)$ as well as $C_{ii}(t)$, the auto-correlation of $i(t)$.

The variance of the conditional state, the residual uncertainty in our knowledge about the system, is deterministic and given by $\Var_c[\hat{X}_\epr](t)\equiv\Var[\hat{X}_{\epr}(t)-\Xeprc(t)]=\Var[\hat{X}_\epr]-\Var[X^c_\epr(t)]$, i.e., the difference between the unconditional (steady-state) variance $\Var[\hat{X}_\epr]$ and the (ensemble) variance of our optimal estimate $\Var[X^c_\epr(t)]=\int_0^t \mathrm{d}t'\, K_{X}(-t',t) C_{Xi}(t')$ 
(see Methods~\ref{app:wiener}). We calculate $\Var[\hat{X}_\epr]$ and $C_{Xi}(t)$ using fitted model parameters.
 In this manner, a complete set of second moments and the full conditional covariance matrix for $(\hat{X}_{\epr},\hat{P}_{\epr})$ are found. The raw experimental photocurrent $i(t)$ is used to obtain the stochastic first moments, fully defining the Gaussian state. $\Var[\hat{X}_\epr]$ contains contributions due to imperfect QBA cancellation and thermal fluctuations, but if the correlation of $\hat{X}_\epr$ with $\Xeprc$ is strong enough, it leads to entanglement as witnessed by $\Var_c[\hat{X}_\epr](t)$. Qualitatively speaking, conditioning suppresses the thermal noise.

We use Wiener filtering to continuously track the EPR oscillator $(\hat{X}_{\epr},\hat{P}_{\epr})$ by inferring the conditional expectation values $\Xeprc\approx (X_\text{M}^c - X_\text{S}^c)/\sqrt{2}$ and $\Peprc\approx (P_\text{M}^c + P_\text{S}^c)/\sqrt{2}$ 
(optimal weights of $\text{M}$ and $\text{S}$ variables are determined by the full model as presented in Methods~\ref{app:entanglement}). 
Demodulating $i(t)$ with $\cos\omega t$ and $\sin\omega t$, respectively, 
we obtain the conditional system quadratures $(\Xeprcrot,\Peprcrot)$, describing the rotating-frame dynamics of the EPR-entangled system (see Fig.~\ref{fig:trajectory}b).
As the conditioning progresses, we obtain a more precise estimate of the conditional system state, as witnessed by the decreasing conditional variance shown in Fig.~\ref{fig:trajectory}c. In the long-conditioning-time limit, maximum information is extracted from past measurements, and the shape of the Wiener filter attains its steady-state form, $K(t'-t,t)\rightarrow K(t'-t)$. 
The corresponding steady-state conditional variance that we observe for a near-resonant case ($\omega_\text{M}\approx-\omega_\text{S}$) is $V_c=\Var_c[\hat{X}_\epr]+\Var_c[\hat{P}_\epr]=\resonantVc<1$, certifying entanglement of the spin and mechanics. This can be directly compared with a case where the frequencies of the systems are not matched, and consequently the best value $V_c=\detunedVc$ is above the entanglement limit. 
In the rest of the paper, we describe the experiment and analysis leading to the entanglement observation.

\begin{figure}[tbp]
    \centering
    \includegraphics[width=\columnwidth]{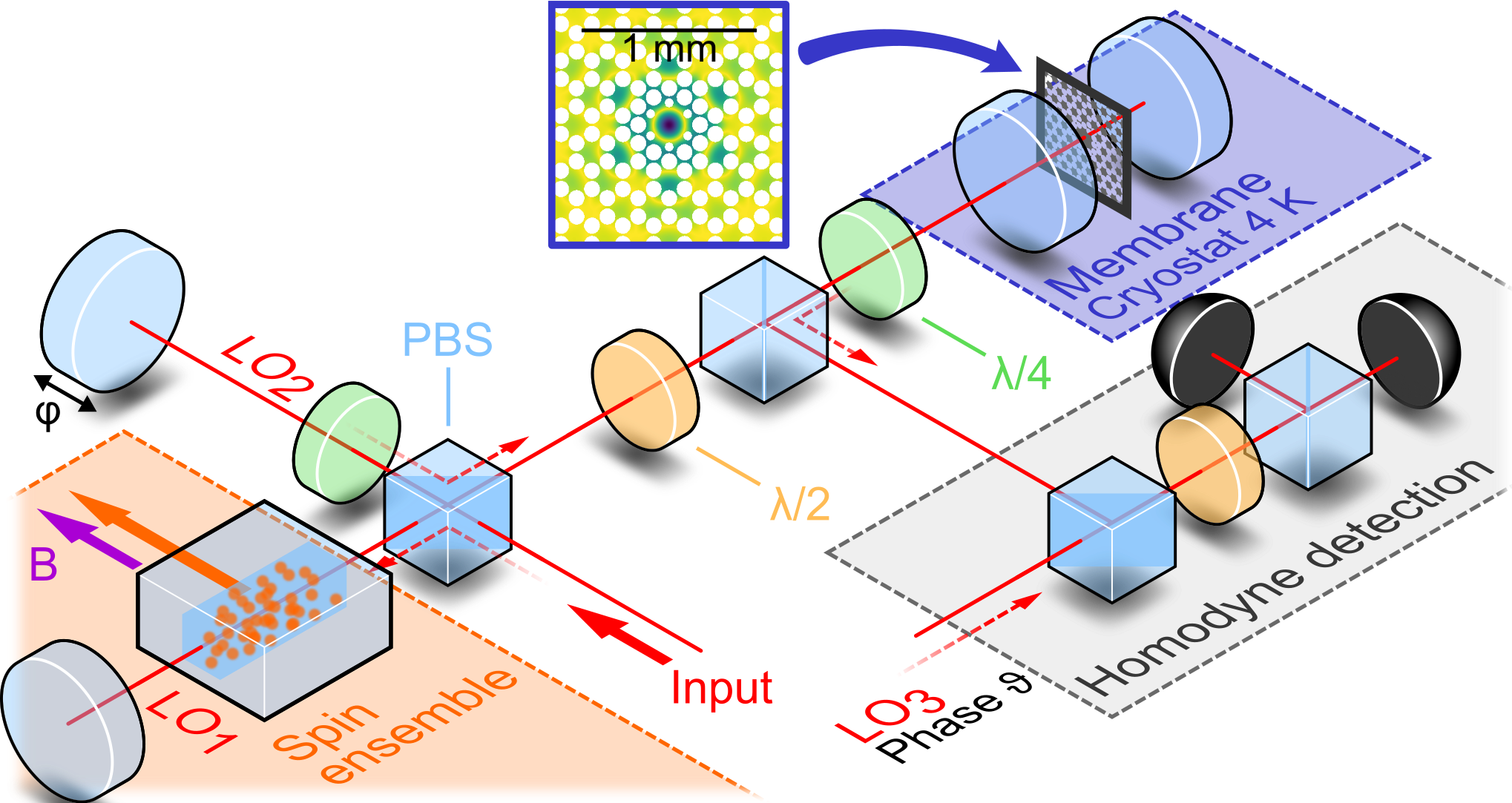}
    \caption{\textbf{Experimental setup for hybrid entanglement generation.} The local oscillator \LO1 reads out the spin system precessing in the magnetic field $B$, with the quantum sideband fields written into the orthogonal light polarisation. After splitting off \LO1, \LO2, phase shifted by $\varphi$ relative to \LO1, is mixed with the sidebands. After projection into a common polarisation, this light is sent to the mechanical system, which is probed in reflection. Final homodyne measurement of the cascaded hybrid system is performed with \LO3, with phase $\vartheta$. $\lambda/2$: Half-wave plate. $\lambda/4$: Quarter-wave plate. PBS: Polarising beam splitter. LO: Local oscillator. See main text for details. Inset: Mode shape of the mechanical mode under investigation (absolute displacement, linear scale).
    }
    \label{fig:expsetup}
\end{figure}

The layout of the hybrid system is outlined in Fig.~\ref{fig:expsetup} (see Methods~\ref{app:exp} for further details). First, the light interacts with the collective spin of a caesium atomic ensemble, contained in a $\SI{300}{\micro\meter}\times\SI{300}{\micro\meter}\times\SI{10}{\milli\meter}$ glass cell. The spin anti-relaxation coating of the cell~\cite{Balabas2010}, along with magnetic shielding, provides a spin coherence lifetime of $T_{2}=\SI{0.7}{\milli\second}$.

Light interacts with the spin ensemble in a double-pass configuration, thus increasing the light-spin interaction strength.
The quantum operators of interest, $\hat X_\text{L}^\text{in}$ and $\hat P_\text{L}^\text{in}$, are the in-phase and in-quadrature vacuum fluctuations of the field polarised orthogonally to \LO1.
Light-matter mapping is well described by the  Hamiltonian $\hat{H}_\textrm{int}/\hbar\propto \sqrt{\Gamma_\textrm{S}}(\hat{X}_\textrm{S}\hat{X}_\textrm{L}+\zeta_\textrm{S}\hat{P}_\textrm{S}\hat{P}_\textrm{L})$, which is close to the quantum non-demolition (QND) interaction as $\zeta_\textrm{S}\sim 0.03$. The interaction leads to a rotation of the input polarisation state~\cite{RevModPhys.82.1041}, with a coupling rate $\Gamma_\text{S}/2\pi=\GammaS$ and the bandwidth $\gamma_\text{S0}/2\pi=\gammaSzero$ (full width at half maximum) due to decoherence processes. The deviation from the QND interaction leads to light partially exchanging states with the oscillator~\cite{Geremia2007,Muschik2011}, additionally broadening the spin oscillator~\cite{Krauter2011} with rate $\delta\gamma_{\text{S}}/2\pi\equiv 2\zeta_\textrm{S}\Gamma_\textrm{S}=\SI{1.2}{\kilo\hertz}$. The spin also couples to its own effective thermal bath with the net stochastic force $\hat{F}_\text{S}$ originating from imperfect optical pumping, spin-exchange collisions, and projection noise, resulting in the mean bath occupation $n_\textrm{S}=\nS$.

The \LO1 is filtered out after interacting with the spin, and the quantum signal is spatially overlapped with \LO2. We mix these two fields into the same polarisation with a waveplate and a polarising beam splitter. This directly translates the polarisation quadrature operators that interacted with the spin system into the amplitude and phase quadratures that are now coupled to the membrane-in-the-middle optomechanical system, in which the radiation pressure of \LO2 drives the mechanical oscillator~\citep{cavityoptomechRMP}. 

The mechanical oscillator is realised in a highly stressed silicon nitride membrane that is \SI{13}{\nano\meter} thick and has millimeter-scale transverse dimensions. The membrane is periodically patterned, leading to the emergence of a phononic bandgap.
The soft-clamped~\citep{softclampingTsaturyan2017} mechanical mode is an out-of-plane, localised center-of-mass vibrational mode~(see inset in Fig.~\ref{fig:expsetup}) with a frequency of $\omega_\text{M0}/2\pi = \SI{\Omegamzero}{\mega\hertz}$ and a quality factor of $Q=650\times 10^6$, i.e., a natural linewidth of $\gamma_{\text{M}0}/2\pi = \SI{2.1}{\milli\hertz}$, at cryogenic operating temperatures. The membrane is placed near the optical beam waist of a \SI{2.6}{\milli\meter} long cavity which has a linewidth of $\kappa/2\pi=\SI{\kappamech}{\mega\hertz}$, and is strongly overcoupled in reflection by \SI{\ovc}{\percent}. 

The optomechanical system is mounted in a \SI{4}{\kelvin} flow cryostat and optically probed. The effective thermal bath at \SI{10}{\kelvin} acts as a stochastic driving force $\hat{F}_\text{M}$ for the mechanical mode of interest. Light is detuned by $\Delta/2\pi \approx \avgDelta$ from the cavity resonance, cooling the mechanical mode to near its motional quantum ground state with a mean phonon occupancy of roughly $2$. This dynamical backaction cooling~\cite{cavityoptomechRMP} broadens the mechanical response to $\gamma_\text{M}/2\pi=\gammaM$ and red-shifts its resonance frequency by \spring\ to $\omega_\text{M}/2\pi=\SI{\Omegam}{\mega\hertz}$. 
The state of the mechanical system is extracted optically at a readout rate of $\Gamma_{\text{M}}/2\pi = \GammaM$. 

Homodyne phase-quadrature measurement of the light reflected off the optomechanical cavity is performed with \LO3. 
The optical transmission between the spin and mechanical systems is $\nu=\effnu$, and the final EPR detection efficiency is $\eta=\effeta$, which includes optical losses in the path between the hybrid system and the detector and the detector quantum efficiency.

In order to construct the Wiener filter and deduce the entanglement from the data we derive the input-output relations for both systems individually and for the hybrid setup.

The response equation for the individual oscillators is $\hat{X}_{j}=\chi_{j}[\hat{F}_{j}+2\sqrt{\Gamma_{j}}(\hat{X}_{\text{L},j}^{\text{in}}\pm i\zeta_{j}\hat{P}_{\text{L},j}^{\text{in}})]$, where $\pm$ signifies $\text{sign}(\omega_{j0})$. The effective, Fourier-domain susceptibility is $\chi_{j(j0)}(\Omega)=\omega_{j0}/(\omega_{j}^{2}-\Omega^2-i\Omega\gamma_{j(j0)})$ including (excluding) the dynamical broadening $\delta\gamma_j \equiv \gamma_j - \gamma_{j0} = 2\zeta_j \Gamma_j$, parametrised in terms of the readout rate $\Gamma_j$ and $1> \zeta_j > -1$. Positive dynamical broadening $\zeta_j>0$ provides beneficial cooling while adding extra QBA noise.

The input-output relation for the optical quadratures $\hat{\bm{X}}_{\text{L},j}^{\text{in(out)}}\equiv (\hat{X}_{\text{L},j}^{\text{in(out)}},\hat{P}_{\text{L},j}^{\text{in(out)}})^\intercal$ probing the individual oscillators is $\hat{\bm{X}}_{\text{L},j}^{\text{out}}=\hat{\bm{X}}_{\text{L},j}^{\text{in}}+\sqrt{\Gamma_{j}}(\pm i\zeta_{j},1)^\intercal\hat{X}_{j}$, showing how $\zeta_j \neq 0$ entails the simultaneous mapping of the oscillator response into both light quadratures (see Methods~\ref{app:model} for details).

In the hybrid experiment light propagates from the spin ensemble to the mechanics, and the phases of the quadratures are adjusted by tuning the phase $\varphi$ between \LO1 and \LO2 (see Fig.~\ref{fig:expsetup}) such that
 $\hat{\bm{X}}_{\text{L,M}}^{\text{in}}=-\sqrt{\nu}\hat{\bm{X}}_{\text{L,S}}^{\text{out}}+\sqrt{1-\nu}\hat{\bm{X}}_{\text{L},\nu}$, where  $\hat{\bm{X}}_{\text{L},\nu}$ is the vacuum field due to intersystem losses.
Whenever $\zeta_{\text{M}}\neq\zeta_{\text{S}}$, a part of the spin response $\hat X_\text{S}$ is mapped into the optical quadrature driving the mechanics, $\hat {X}_{\text{L,M}}^\text{in} + i\zeta_\text{M}\hat{P}_\text{L,M}^\text{in}$; this enables non-local dynamical cooling of the combined EPR oscillator, a mechanism related to unconditional entanglement generation~\citep{Krauter2011, huangzeuthenprl2018}. The EPR readout impinging on the detector is $\hat{\bm{X}}_{\text{L}}^{\text{out}}=\sqrt{\eta}\hat{\bm{X}}_{\text{L,M}}^{\text{out}}+\sqrt{1-\eta}\hat{\bm{X}}_{\text{L},\eta}$, accounting for the finite EPR detection efficiency $\eta$. 

Combining those relations,  we obtain the EPR readout through the phase quadrature of light
\begin{subequations}\label{eq:IO_intro}
\begin{align}
&\hat{P}_{\text{L}}^{\text{out}}=\hat{P}_{\text{L}}^{\text{in}\prime} + \sqrt{\eta}\left(\sqrt{\Gamma_{\text{M}}}\hat{X}_{\text{M}}-\sqrt{\nu\Gamma_{\text{S}}}\hat{X}_{\text{S}}\right)\label{eq:IO_intro-a}\\
&\approx \hat{P}_{\text{L}}^{\text{in}\prime}\!+\!\sqrt{\eta}\bigg(\!-\!\sqrt{\nu}\left[\tfrac{\chi_{\text{S}}}{\chi_{\text{S}0}}\Gamma_{\text{M}}\chi_{\text{M}}+\tfrac{\chi_{\text{M}}}{\chi_{\text{M}0}}\Gamma_{\text{S}}\chi_{\text{S}}\right]2\hat{X}_{\text{L,S}}^{\text{in}}\label{eq:IO_intro-b}\\ 
& +\sqrt{\Gamma_{\text{M}}}\chi_{\text{M}}[\hat{F}_{\text{M}}+\sqrt{(1-\nu)\Gamma_{\text{M}}}2\hat{X}_{\text{L},\nu}]-\tfrac{\chi_{\text{M}}}{\chi_{\text{MS}}}\sqrt{\nu\Gamma_{\text{S}}}\chi_{\text{S}}\hat{F}_{\text{S}}\bigg),\nonumber
\end{align}
\end{subequations}
where $\hat{P}_{\text{L}}^{\text{in}\prime}$ is the measurement imprecision noise including shot noise and broadband spin noise due to imperfect motional averaging~\cite{Borregaard2016}. 
The second line of Eq.~(\ref{eq:IO_intro-b}) contains the uncorrelated noise contributions driving the individual subsystems: intrinsic thermal and ground state noise $\hat{F}_{j}$ and the extraneous QBA $\hat{X}_{\text{L},\nu}$. 

The thermal forces acting on subsystem $j$ are suppressed due to the dynamical cooling $\delta\gamma_j >0$ (contained in $\chi_j$) occurring locally at each subsystem. Additionally, the thermal spin response $\chi_{\text{S}}\hat{F}_{\text{S}}$ is further suppressed by $\chi_{\text{M}}/\chi_{\text{MS}}$ due to the non-local dynamical EPR cooling, introducing the cross-susceptibility $\chi_{\text{MS}}^{-1}(\Omega)\equiv\chi_{\text{M}0}^{-1}(\Omega)-i2\zeta_{\text{S}}\Gamma_{\text{M}}$.

The joint QBA term $\propto\hat{X}_{\text{L,S}}^{\text{in}}$ in the first line of Eq.~(\ref{eq:IO_intro-b}) embodies the central physical mechanism of our scheme resulting from the following two interfering processes:  
first, the spin system produces squeezed amplitude fluctuations $\hat{X}_{\text{L,S}}^{\text{out}}\sim(\chi_{\text{S}}/\chi_{\text{S}0})\hat{X}_{\text{L,S}}^{\text{in}}$ which map into the mechanical phase quadrature response according to $\Gamma_{\text{M}}\chi_{\text{M}}$; second, the spin QBA response $\hat{P}_{\text{L,S}}^{\text{out}}\sim\Gamma_{\text{S}}\chi_{\text{S}}\hat{X}_{\text{L,S}}^{\text{in}}$ is subsequently filtered by the mechanical system according to $\chi_{\text{M}}/\chi_{\text{M}0}$. We remark that the function $\chi_{j}/\chi_{j0}$ suppresses near-resonant spectral components in a bandwidth $\gamma_j$ with maximal suppression $\gamma_{j0}/\gamma_j$ at $\Omega\sim\omega_j$ (for $\delta\gamma_j >0$). Since $\gamma_{\text{M}0}/\gamma_{\text{M}} \ll 1$, this entails strong suppression of the spin QBA response, whereas the amplitude squeezing by the spin is more moderate $\gamma_{\text{S}0}/\gamma_{\text{S}}\approx \gammaSratio$. 

The prefactor to $\hat X_{\text{L,S}}^\text{in}$ may be rewritten as $\chi_\text{M}\chi_\text{S}/(\chi_{\text{M}0}\chi_{\text{S}0})[\Gamma_\text{M}\chi_{\text{M}0} + \Gamma_\text{S}\chi_{\text{S}0}]$, highlighting the condition $\Gamma_{\text{M}}\chi_{\text{M}0}+\Gamma_{\text{S}}\chi_{\text{S}0}=0$ for total broadband BA cancellation (independent of dynamical broadening), which requires $\omega_\text{M} = -\omega_\text{S}$. In the case of unmatched intrinsic linewidths  $\gamma_{\text{M}0}\neq\gamma_{\text{S}0}$, one still needs $\omega_\text{M} = -\omega_\text{S}$ to minimise the term. 
Eq.~(\ref{eq:IO_intro-b}) demonstrates that dynamical broadening enhances QBA suppression significantly via the factors $\chi_j/\chi_{j0}$.
  
While BA reduction is necessary to achieve $V<1$, it is not sufficient, due to the inevitable presence of ground-state fluctuations contained in $\hat{F}_j$, cf.\ Eq.~(\ref{eq:IO_intro-b}). 
These thermal fluctuations, along with residual QBA, can be suppressed by the conditional tracking and/or the coherent dynamical cooling mechanisms (local and non-local) discussed above; here we simultaneously employ both types of mechanisms.

While the model~(\ref{eq:IO_intro}) captures all essential aspects of the involved EPR dynamics, certain technical or peripheral effects were left out for simplicity. 
These include the finite overcoupling of the optical cavity and the option of introducing optical quadrature rotations between the subsystems as well as in the homodyne detection.
Moreover, Eq.~(\ref{eq:IO_intro-b}) neglects the phase noise BA contributions $\propto \hat{P}_{\text{L,S}}^{\text{in}},\hat{P}_{\text{L},\nu}$ to the EPR response, which are minor for the parameter regime considered here. 
The full model accounting for all aforementioned effects was employed in analysing the experimental data (see Methods~\ref{app:model}).

\begin{figure}[tbp]
    \centering
    \includegraphics[width=\columnwidth]{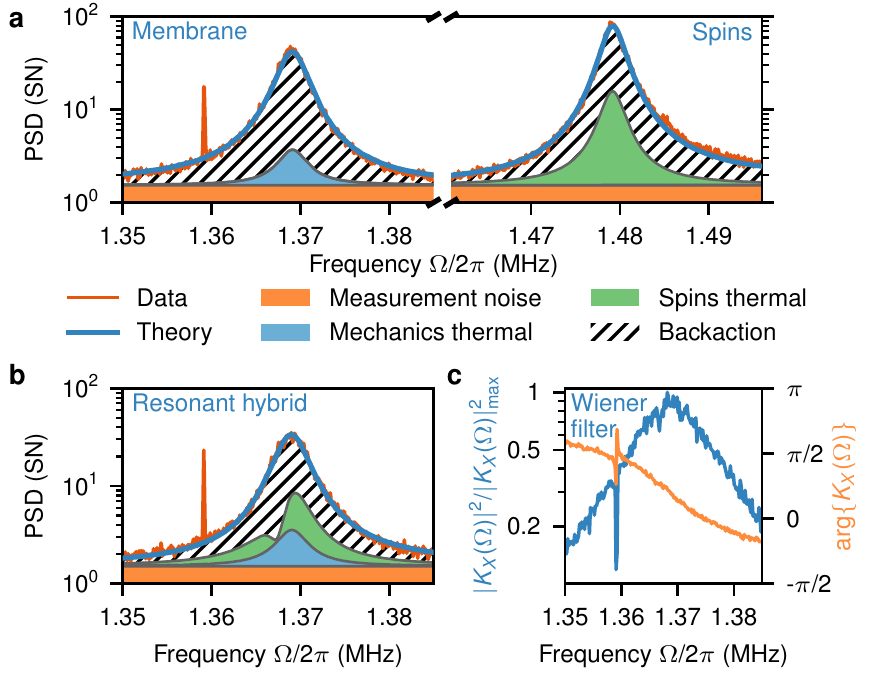}
    \caption{\textbf{Quantum noise spectra of the hybrid system}.  \textbf{a,}  Optical phase quadrature power spectral densities (PSD) for the measurements of the individual oscillators (detuned by 110 kHz). The feature at $\sim\SI{1.359}{\MHz}$ is due to laser phase noise. \textbf{b,} Joint spectrum of the EPR system for the spin oscillator tuned close to resonance with the mechanical oscillator. Notably, the relative (as well as absolute) amount of QBA noise in the joint signal is significantly reduced compared to that of the individual oscillators. \textbf{c,} Wiener filter normalized absolute square amplitude (blue, left axis) and phase (orange, right axis) for the resonant hybrid case. The filtering procedure discerns the hybrid system signal from experimental imperfections, e.g., the laser phase noise peak.
    }
    \label{fig:spectra}
\end{figure}

The various noise suppression mechanisms of \eqref{IO_intro-b} manifest themselves in the noise spectra of the optical readout $\hat P_\text{L}^\text{out}$ (see Fig.~\ref{fig:spectra}). 
When the two oscillators are detuned by \SI{110}{\kilo\hertz}, by changing $\omega_{\text{S}}\propto B$, they are essentially probed separately~(Fig.~\ref{fig:spectra}a).  
However, due to the finite detuning, the measurement noise for both subsystems is the sum of optical shot noise and broadband spin noise. 
The Lorentzian features $\propto |\chi_j|^2$ are the dynamically broadened responses to thermal noise and light QBA of the two systems. The mechanical motion is strongly dominated by QBA, its ratio to intrinsic thermal noise (TH) being $\text{QBA}/\text{TH}=\QBATHM$, whereas for the spin oscillator $\text{QBA}/\text{TH}=\QBATHS$. 

When the spin oscillator is tuned close to the mechanical resonance~(Fig.~\ref{fig:spectra}b), we observe strong overall noise reduction for the EPR oscillator. 
Firstly, the non-local dynamical cooling of the spin thermal noise amounts to a reduction of the joint noise due to stochastic forces $\hat{F}_\text{M}$ and $\hat{F}_\text{S}$ by $\jointThermalReduction$. Secondly, we observe the reduction of the QBA due to the destructive interference 
by \resonantBAreduction (striped area in Fig.~\ref{fig:spectra}b), compared to the sum of QBAs of the two separate systems (striped areas in Fig.~\ref{fig:spectra}a). As a result, the unconditional EPR variance is reduced by $\SI{\resonantVureduction}{\decibel}$, from $\detunedVune$ to $\resonantVune$, as has already been indicated in Fig.~\ref{fig:trajectory}c. 
The asymmetrical features in Fig.~\ref{fig:spectra}b arise due to the small but finite spin-mechanics detuning and the choice of the \LO2 phase $\varphi$.

Having discussed the coherent suppression of QBA and thermal noise, which reduces the unconditional EPR variance $\Vu$ in our experiment, we now return to the Wiener filtering method we apply to verify conditional entanglement $\Vc < 1$. To focus on its essential properties, we here consider the QND limit $\gamma_j = \gamma_{j0}$ of \eqref{IO_intro}. 

Before considering the hybrid system, we apply the filtering method to the tracking of a single oscillator, specifically the mechanical one, by setting $\Gamma_\text{S}=0$ (and $\Gamma_{\text{M}}\equiv\Gamma$). The same argument can be made for the tracking of the spin oscillator.
In this case, the filtering reduces the variance from its unconditional value $V_\text{u}^{(1)}=(1+2n)(1+\Cq)$ 
to 
$\Vc^{(1)} \approx \sqrt{1/(2\eta)} \sqrt{1+(\gamma/\Gamma)\Vu^{(1)}} = \sqrt{1/\eta} \sqrt{1+1/(2 \Cq)}$, assuming the rotating wave approximation ($\omega \gg \gamma\Vu^{(1)}, \Gamma$) and fast readout $\sqrt{8\eta\Vu^{(1)}\Gamma/\gamma}\gg 1$ here and henceforth~\citep{LammersThesis}. 
We have expressed the quantum cooperativity $\Cq=\Gamma/(\gamma[2 n +1])$ in terms of the thermal bath occupancy $n$. Within the QND model, $\Cq = \textrm{QBA}/\textrm{TH}$. An efficient measurement of a single system, $\eta=1$ and $\Cq\rightarrow\infty$, can bring the conditional variance to the ground-state value $\Vc^{(1)}\rightarrow 1$, but not below.  

In the idealised hybrid case with matched readout rates $\Gamma_{\text{M}}=\Gamma_{\text{S}}\equiv \Gamma$ and intrinsic susceptibilities $\chi_{\text{M}0} = -\chi_{\text{S}0}$ (implying $\omega_{\text{M}0} = -\omega_{\text{S}0} \equiv \omega$ and  $\gamma_{\text{M}0} = \gamma_{\text{S}0} \equiv \gamma$), 
the tracking reduces the variance from the QBA-free unconditional value $\Vu=1+2n$ (where now $n=(n_\text{S}+n_\text{M})/2$) to 
\begin{equation}\label{eq:Vc-epr}
     \Vc \approx \frac{1}{\sqrt{2\eta}}\sqrt{\frac{\gamma}{\Gamma}\Vu} = \frac{1}{\sqrt{\eta}}\sqrt{\frac{1}{2 \Cq}},
\end{equation}
which shows that ideal QBA cancellation removes the lower bound $\Vc^{(1)}\geq 1$ associated with the single-oscillator case.

When applied to experimental data, the Wiener filter not only optimally discerns the EPR signal from white measurement noise, but also rejects other coloured, peaked, or cross-correlated noises. 
Fig.~\ref{fig:spectra}c presents the steady-state, frequency-domain Wiener filter $K_{X}(\Omega)\propto \int_{-\infty}^\infty\mathrm{d}\tau\, \exp(i\Omega \tau)K_{X}(\tau)$ for the hybrid case of nearly resonant oscillators. It takes into account the full model of the joint system along with experimentally measured noise.

\begin{figure}[t]
    \centering
    \includegraphics[]{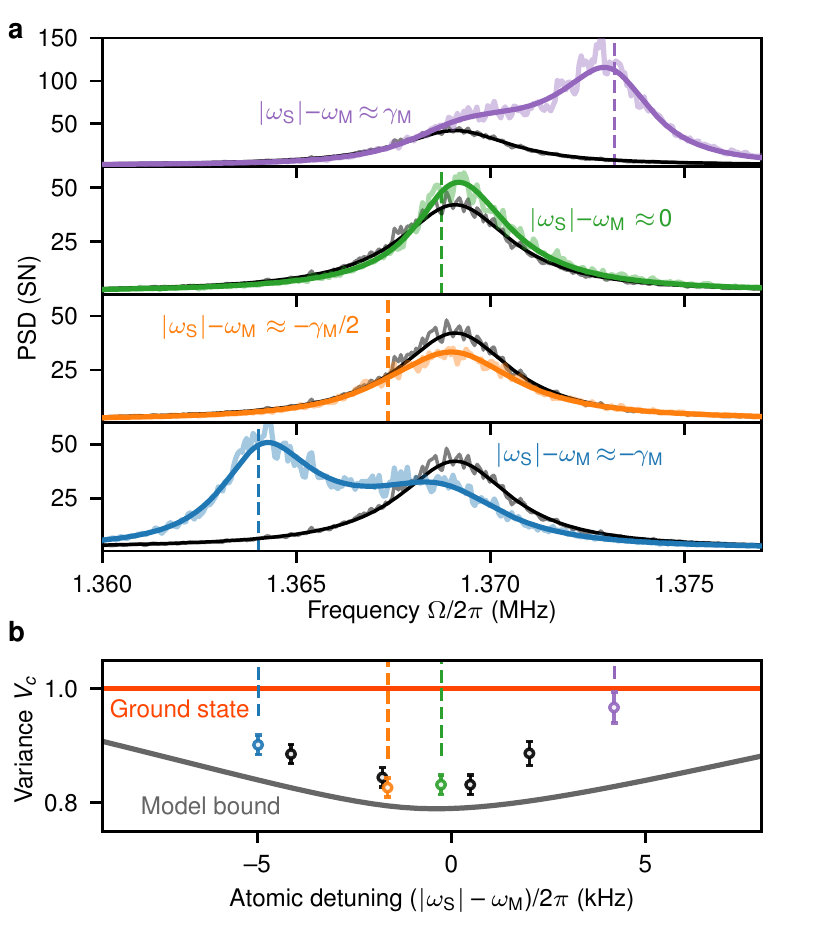}
    \caption{\textbf{Entanglement tuning and optimization}. We sweep the resonant (Larmor) frequency of the spin oscillator across the mechanical resonance. This tunes the distinguishability of the two oscillators in the measured light, thus varying the achievable conditional EPR entanglement. \textbf{a} Spectra with respective fits, and \textbf{b} resulting variance for different values of the mechanical-spin detuning. Black curves in \textbf{a} depict the mechanics only spectrum. Dashed vertical lines indicate spin frequency. Black curve in \textbf{b} depict variances of optimized EPR combinations, calculated from theoretical spectra, given parameter values from the orange point. Excess experimental noise leads to expected deviations of this theoretical bound and experimental values. For the spectra corresponding to the black points see Methods, Fig.~\ref{fig:histograms}.
    }
    \label{fig:entanglement}
\end{figure}

In the experiment, the conditional variance is determined by many factors, such as optical losses $\nu$ and $\eta$, as well as mismatched intrinsic linewidths $\gamma_{\text{M}0}\neq\gamma_{\text{S}0}$ and readout rates $\Gamma_j$. 
Fig.~\ref{fig:spectra} presents results for the case  $\Gamma_{\text{M}}\approx\nu\Gamma_{\text{S}}$.
Whereas in Fig.~\ref{fig:spectra}b we have matched frequencies $\omega_\text{M}\approx -\omega_\text{S}$, in Fig.~\ref{fig:entanglement}a we present a series of spectra in which $|\omega_{\text{S}}|$ is swept through the mechanical resonance by tuning the $B$ field. 
The resulting \Vc for a set of such measurements~(Fig.~\ref{fig:entanglement}b) exhibits a smooth transition between the regimes of entangled and non-entangled states of the hybrid system. 
Notably, our system is rather resilient even to quite substantial oscillator detunings $|\omega_{\text{S}}|-\omega_{\text{M}}$. The bandwidth of $V_c$ as a function of the detuning is affected by the readout rates, here amounting to several oscillator linewidths. 

Using the entire set of measurements we evaluate the uncertainty of the final result for $\Vc$ as discussed in Methods~\ref{app:mcmc}. In short, we establish priors for all experimental parameters by independent measurements, and use Markov Chain Monte Carlo simulations and log-likelihood optimisation to obtain fit parameters and their uncertainties, in particular obtaining the uncertainty for the conditional EPR variance.

In conclusion, we have demonstrated entanglement between distant objects in a hybrid system consisting of a mechanical oscillator and an atomic spin ensemble.
This constitutes a new milestone for hybrid macroscopic entanglement and for demonstration of noiseless trajectories in the negative-mass reference frame. 

Future applications of the noiseless measurement of motion under realistic thermal conditions enabled by this work include, e.g., off-resonant continuous force detection~\cite{KhaliliESPprl2018,Zeuthenpolzikhaliliprd2019} and resonant pulsed measurements based on state preparation and retrodiction~\cite{Wasilewski2010}.

Future work on enhancing entanglement and achieving practical detection of noiseless trajectories of motion will primarily concentrate on amending experimental imperfections. This includes reduction of broadband spin noise by better mode-matching of light to the atomic ensemble, reducing optical losses and the cavity noise that appears from the mirror-modes, as well as motional drifts due to thermal fluctuations of the cryogenic chamber. 
Reduction of intersystem optical losses down to 10\%, improvement of cavity overcoupling to $\kappa_\text{in}/\kappa=0.98$, improvement of the fractional coherent spin readout $\Gamma_{\text{S}}/\gamma_{\text{S}0}$ by a factor of $3$, and reduction of spin broadband noise by a factor of $3$ will lead to $\Vc \approx 0.3$ ($\SI{-5}{\decibel}$) according to our model.

\begin{acknowledgements}
The authors acknowledge enlightening conversations with K.\,Hammerer and J.\,H.\,Müller. X.\,Huang contributed theoretical simulations in the early stages of the project. We acknowledge M.\,Balabas for fabricating the coated caesium cells.

This project has been supported by the European Research Council Advanced grant QUANTUM-N, the Villum Foundation, and John Templeton Foundation. E.\,Z.\ acknowledges funding from the Carlsberg Foundation. 
 M.\,P.\ was partially supported by the Foundation for Polish Science (FNP). R.\,A.\,T.\ was partially funded by the program Science without Borders of the Brazilian Federal Government.  
\end{acknowledgements}

E.\,S.\,P.\ conceived and led the project. 
R.\,A.\,T., C.\,B.\, M., C.\,Ø.\ and M.\,P.\ built the experiment with the help of C.\,B.\ and J.\,A.. 
The membrane resonator was conceived by A.\,S.\ and was designed and fabricated by Y.\,T..
R.\,A.\,T., M.\,P., C.\,Ø., C.\,B.\,M.\ and C.\,B.\ took the data. 
E.\,Z.\ and M.\,P.\ developed the theory with input from J.\,A., E.\,S.\,P. and R.\,A.\,T.. 
The paper was written by E.\,S.\,P., E.\,Z., R.\,A.\,T., M.\,P., C.\,Ø., C.\,B.\,M.\  and C.\,B., with contributions from other authors.

\bibliographystyle{naturemag}
\bibliography{references.bib}

\begin{thebibliography}{10}
\expandafter\ifx\csname url\endcsname\relax
  \def\url#1{\texttt{#1}}\fi
\expandafter\ifx\csname urlprefix\endcsname\relax\def\urlprefix{URL }\fi
\providecommand{\bibinfo}[2]{#2}
\providecommand{\eprint}[2][]{\url{#2}}

\bibitem{KimbleNature2008}
\bibinfo{author}{Kimble, H.~J.}
\newblock \bibinfo{title}{The quantum internet}.
\newblock \emph{\bibinfo{journal}{Nature}} \textbf{\bibinfo{volume}{453}},
  \bibinfo{pages}{1023--1030} (\bibinfo{year}{2008}).

\bibitem{Kurizki2015}
\bibinfo{author}{Kurizki, G.} \emph{et~al.}
\newblock \bibinfo{title}{Quantum technologies with hybrid systems}.
\newblock \emph{\bibinfo{journal}{Proceedings of the National Academy of
  Sciences}} \textbf{\bibinfo{volume}{112}}, \bibinfo{pages}{3866--3873}
  (\bibinfo{year}{2015}).

\bibitem{Degen2017}
\bibinfo{author}{Degen, C.~L.}, \bibinfo{author}{Reinhard, F.} \&
  \bibinfo{author}{Cappellaro, P.}
\newblock \bibinfo{title}{Quantum sensing}.
\newblock \emph{\bibinfo{journal}{Rev. Mod. Phys.}}
  \textbf{\bibinfo{volume}{89}}, \bibinfo{pages}{035002}
  (\bibinfo{year}{2017}).

\bibitem{Chen2013}
\bibinfo{author}{Chen, Y.}
\newblock \bibinfo{title}{Macroscopic quantum mechanics: theory and
  experimental concepts of optomechanics}.
\newblock \emph{\bibinfo{journal}{Journal of Physics B: Atomic, Molecular and
  Optical Physics}} \textbf{\bibinfo{volume}{46}}, \bibinfo{pages}{104001}
  (\bibinfo{year}{2013}).

\bibitem{cavityoptomechRMP}
\bibinfo{author}{Aspelmeyer, M.}, \bibinfo{author}{Kippenberg, T.~J.} \&
  \bibinfo{author}{Marquardt, F.}
\newblock \bibinfo{title}{Cavity optomechanics}.
\newblock \emph{\bibinfo{journal}{Rev. Mod. Phys.}}
  \textbf{\bibinfo{volume}{86}}, \bibinfo{pages}{1391--1452}
  (\bibinfo{year}{2014}).

\bibitem{Duan2000}
\bibinfo{author}{Duan, L.-M.}, \bibinfo{author}{Giedke, G.},
  \bibinfo{author}{Cirac, J.~I.} \& \bibinfo{author}{Zoller, P.}
\newblock \bibinfo{title}{Inseparability criterion for continuous variable
  systems}.
\newblock \emph{\bibinfo{journal}{Phys. Rev. Lett.}}
  \textbf{\bibinfo{volume}{84}}, \bibinfo{pages}{2722--2725}
  (\bibinfo{year}{2000}).

\bibitem{softclampingTsaturyan2017}
\bibinfo{author}{Tsaturyan, Y.}, \bibinfo{author}{Barg, A.},
  \bibinfo{author}{Polzik, E.~S.} \& \bibinfo{author}{Schliesser, A.}
\newblock \bibinfo{title}{Ultracoherent nanomechanical resonators via soft
  clamping and dissipation dilution}.
\newblock \emph{\bibinfo{journal}{Nature Nanotechnology}}
  \textbf{\bibinfo{volume}{12}}, \bibinfo{pages}{776--783}
  (\bibinfo{year}{2017}).

\bibitem{Borregaard2016}
\bibinfo{author}{Borregaard, J.} \emph{et~al.}
\newblock \bibinfo{title}{Scalable photonic network architecture based on
  motional averaging in room temperature gas}.
\newblock \emph{\bibinfo{journal}{Nature Communications}}
  \textbf{\bibinfo{volume}{7}}, \bibinfo{pages}{11356} (\bibinfo{year}{2016}).

\bibitem{julsgaard2001experimental}
\bibinfo{author}{Julsgaard, B.}, \bibinfo{author}{Kozhekin, A.} \&
  \bibinfo{author}{Polzik, E.~S.}
\newblock \bibinfo{title}{Experimental long-lived entanglement of two
  macroscopic objects}.
\newblock \emph{\bibinfo{journal}{Nature}} \textbf{\bibinfo{volume}{413}},
  \bibinfo{pages}{400} (\bibinfo{year}{2001}).

\bibitem{eprpolzikhammerer2009}
\bibinfo{author}{Hammerer, K.}, \bibinfo{author}{Aspelmeyer, M.},
  \bibinfo{author}{Polzik, E.~S.} \& \bibinfo{author}{Zoller, P.}
\newblock \bibinfo{title}{Establishing {E}instein-{P}oldosky-{R}osen channels
  between nanomechanics and atomic ensembles}.
\newblock \emph{\bibinfo{journal}{Phys. Rev. Lett.}}
  \textbf{\bibinfo{volume}{102}}, \bibinfo{pages}{020501}
  (\bibinfo{year}{2009}).

\bibitem{polzikadp2015}
\bibinfo{author}{Polzik, E.~S.} \& \bibinfo{author}{Hammerer, K.}
\newblock \bibinfo{title}{Trajectories without quantum uncertainties}.
\newblock \emph{\bibinfo{journal}{Annalen der Physik}}
  \textbf{\bibinfo{volume}{527}}, \bibinfo{pages}{A15--A20}
  (\bibinfo{year}{2015}).

\bibitem{Moller2017}
\bibinfo{author}{M{\o}ller, C.~B.} \emph{et~al.}
\newblock \bibinfo{title}{Quantum back-action-evading measurement of motion in
  a negative mass reference frame}.
\newblock \emph{\bibinfo{journal}{Nature}} \textbf{\bibinfo{volume}{547}},
  \bibinfo{pages}{191--195} (\bibinfo{year}{2017}).

\bibitem{tsangcavesprl2010}
\bibinfo{author}{Tsang, M.} \& \bibinfo{author}{Caves, C.~M.}
\newblock \bibinfo{title}{Coherent quantum-noise cancellation for
  optomechanical sensors}.
\newblock \emph{\bibinfo{journal}{Phys. Rev. Lett.}}
  \textbf{\bibinfo{volume}{105}}, \bibinfo{pages}{123601}
  (\bibinfo{year}{2010}).

\bibitem{KhaliliESPprl2018}
\bibinfo{author}{Khalili, F.~Y.} \& \bibinfo{author}{Polzik, E.~S.}
\newblock \bibinfo{title}{Overcoming the standard quantum limit in
  gravitational wave detectors using spin systems with a negative effective
  mass}.
\newblock \emph{\bibinfo{journal}{Phys. Rev. Lett.}}
  \textbf{\bibinfo{volume}{121}}, \bibinfo{pages}{031101}
  (\bibinfo{year}{2018}).

\bibitem{Zeuthenpolzikhaliliprd2019}
\bibinfo{author}{Zeuthen, E.}, \bibinfo{author}{Polzik, E.~S.} \&
  \bibinfo{author}{Khalili, F.~Y.}
\newblock \bibinfo{title}{Gravitational wave detection beyond the standard
  quantum limit using a negative-mass spin system and virtual rigidity}.
\newblock \emph{\bibinfo{journal}{Phys. Rev. D}}
  \textbf{\bibinfo{volume}{100}}, \bibinfo{pages}{062004}
  (\bibinfo{year}{2019}).

\bibitem{RMPentanglement2009}
\bibinfo{author}{Horodecki, R.}, \bibinfo{author}{Horodecki, P.},
  \bibinfo{author}{Horodecki, M.} \& \bibinfo{author}{Horodecki, K.}
\newblock \bibinfo{title}{Quantum entanglement}.
\newblock \emph{\bibinfo{journal}{Rev. Mod. Phys.}}
  \textbf{\bibinfo{volume}{81}}, \bibinfo{pages}{865--942}
  (\bibinfo{year}{2009}).

\bibitem{CiracZoller95}
\bibinfo{author}{Cirac, J.~I.} \& \bibinfo{author}{Zoller, P.}
\newblock \bibinfo{title}{Quantum computations with cold trapped ions}.
\newblock \emph{\bibinfo{journal}{Phys. Rev. Lett.}}
  \textbf{\bibinfo{volume}{74}}, \bibinfo{pages}{4091--4094}
  (\bibinfo{year}{1995}).

\bibitem{GrossBlochScience2017}
\bibinfo{author}{Gross, C.} \& \bibinfo{author}{Bloch, I.}
\newblock \bibinfo{title}{{Quantum simulations with ultracold atoms in optical
  lattices}}.
\newblock \emph{\bibinfo{journal}{Science}} \textbf{\bibinfo{volume}{357}},
  \bibinfo{pages}{995--1001} (\bibinfo{year}{2017}).

\bibitem{BrownKimMonroenpjQI2016}
\bibinfo{author}{Brown, K.~R.}, \bibinfo{author}{Kim, J.} \&
  \bibinfo{author}{Monroe, C.}
\newblock \bibinfo{title}{{Co-designing a scalable quantum computer with
  trapped atomic ions}}.
\newblock \emph{\bibinfo{journal}{{npj Quantum Information}}}
  \textbf{\bibinfo{volume}{2}} (\bibinfo{year}{2016}).

\bibitem{RevModPhys.82.1041}
\bibinfo{author}{Hammerer, K.}, \bibinfo{author}{S\o{}rensen, A.~S.} \&
  \bibinfo{author}{Polzik, E.~S.}
\newblock \bibinfo{title}{Quantum interface between light and atomic
  ensembles}.
\newblock \emph{\bibinfo{journal}{Rev. Mod. Phys.}}
  \textbf{\bibinfo{volume}{82}}, \bibinfo{pages}{1041--1093}
  (\bibinfo{year}{2010}).

\bibitem{Muschik2011}
\bibinfo{author}{Muschik, C.~A.}, \bibinfo{author}{Polzik, E.~S.} \&
  \bibinfo{author}{Cirac, J.~I.}
\newblock \bibinfo{title}{Dissipatively driven entanglement of two macroscopic
  atomic ensembles}.
\newblock \emph{\bibinfo{journal}{Phys. Rev. A}} \textbf{\bibinfo{volume}{83}},
  \bibinfo{pages}{052312} (\bibinfo{year}{2011}).

\bibitem{Krauter2011}
\bibinfo{author}{Krauter, H.} \emph{et~al.}
\newblock \bibinfo{title}{Entanglement generated by dissipation and steady
  state entanglement of two macroscopic objects}.
\newblock \emph{\bibinfo{journal}{Phys. Rev. Lett.}}
  \textbf{\bibinfo{volume}{107}}, \bibinfo{pages}{080503}
  (\bibinfo{year}{2011}).

\bibitem{Stannigel2012}
\bibinfo{author}{Stannigel, K.}, \bibinfo{author}{Rabl, P.} \&
  \bibinfo{author}{Zoller, P.}
\newblock \bibinfo{title}{Driven-dissipative preparation of entangled states in
  cascaded quantum-optical networks}.
\newblock \emph{\bibinfo{journal}{New Journal of Physics}}
  \textbf{\bibinfo{volume}{14}}, \bibinfo{pages}{063014}
  (\bibinfo{year}{2012}).

\bibitem{Vasilyev2013}
\bibinfo{author}{Vasilyev, D.~V.}, \bibinfo{author}{Muschik, C.~A.} \&
  \bibinfo{author}{Hammerer, K.}
\newblock \bibinfo{title}{Dissipative versus conditional generation of gaussian
  entanglement and spin squeezing}.
\newblock \emph{\bibinfo{journal}{Phys. Rev. A}} \textbf{\bibinfo{volume}{87}},
  \bibinfo{pages}{053820} (\bibinfo{year}{2013}).

\bibitem{huangzeuthenprl2018}
\bibinfo{author}{Huang, X.} \emph{et~al.}
\newblock \bibinfo{title}{Unconditional steady-state entanglement in
  macroscopic hybrid systems by coherent noise cancellation}.
\newblock \emph{\bibinfo{journal}{Phys. Rev. Lett.}}
  \textbf{\bibinfo{volume}{121}}, \bibinfo{pages}{103602}
  (\bibinfo{year}{2018}).

\bibitem{Manukhova2020}
\bibinfo{author}{Manukhova, A.~D.}, \bibinfo{author}{Rakhubovsky, A.~A.} \&
  \bibinfo{author}{Filip, R.}
\newblock \bibinfo{title}{{Pulsed atom-mechanical quantum non-demolition
  gate}}.
\newblock \emph{\bibinfo{journal}{{npj Quantum Information}}}
  \textbf{\bibinfo{volume}{6}}, \bibinfo{pages}{1--8} (\bibinfo{year}{2020}).

\bibitem{cavestsangprx2012}
\bibinfo{author}{Tsang, M.} \& \bibinfo{author}{Caves, C.~M.}
\newblock \bibinfo{title}{Evading quantum mechanics: Engineering a classical
  subsystem within a quantum environment}.
\newblock \emph{\bibinfo{journal}{Phys. Rev. X}} \textbf{\bibinfo{volume}{2}},
  \bibinfo{pages}{031016} (\bibinfo{year}{2012}).

\bibitem{stamperkurn-2018}
\bibinfo{author}{Kohler, J.}, \bibinfo{author}{Gerber, J.~A.},
  \bibinfo{author}{Dowd, E.} \& \bibinfo{author}{Stamper-Kurn, D.~M.}
\newblock \bibinfo{title}{Negative-mass instability of the spin and motion of
  an atomic gas driven by optical cavity backaction}.
\newblock \emph{\bibinfo{journal}{Phys. Rev. Lett.}}
  \textbf{\bibinfo{volume}{120}}, \bibinfo{pages}{013601}
  (\bibinfo{year}{2018}).

\bibitem{jockel2015sympathetic}
\bibinfo{author}{J{\"o}ckel, A.} \emph{et~al.}
\newblock \bibinfo{title}{Sympathetic cooling of a membrane oscillator in a
  hybrid mechanical--atomic system}.
\newblock \emph{\bibinfo{journal}{Nature Nanotechnology}}
  \textbf{\bibinfo{volume}{10}}, \bibinfo{pages}{55} (\bibinfo{year}{2015}).

\bibitem{christoph2018combined}
\bibinfo{author}{Christoph, P.} \emph{et~al.}
\newblock \bibinfo{title}{Combined feedback and sympathetic cooling of a
  mechanical oscillator coupled to ultracold atoms}.
\newblock \emph{\bibinfo{journal}{New Journal of Physics}}
  \textbf{\bibinfo{volume}{20}}, \bibinfo{pages}{093020}
  (\bibinfo{year}{2018}).

\bibitem{Tan2013}
\bibinfo{author}{Tan, H.}, \bibinfo{author}{Buchmann, L.~F.},
  \bibinfo{author}{Seok, H.} \& \bibinfo{author}{Li, G.}
\newblock \bibinfo{title}{Achieving steady-state entanglement of remote
  micromechanical oscillators by cascaded cavity coupling}.
\newblock \emph{\bibinfo{journal}{Phys. Rev. A}} \textbf{\bibinfo{volume}{87}},
  \bibinfo{pages}{022318} (\bibinfo{year}{2013}).

\bibitem{WoolleyClerkPRA2013}
\bibinfo{author}{Woolley, M.~J.} \& \bibinfo{author}{Clerk, A.~A.}
\newblock \bibinfo{title}{Two-mode back-action-evading measurements in cavity
  optomechanics}.
\newblock \emph{\bibinfo{journal}{Phys. Rev. A}} \textbf{\bibinfo{volume}{87}},
  \bibinfo{pages}{063846} (\bibinfo{year}{2013}).

\bibitem{Ockeloen-KorppiNature2018}
\bibinfo{author}{Ockeloen-Korppi, C.~F.} \emph{et~al.}
\newblock \bibinfo{title}{Stabilized entanglement of massive mechanical
  oscillators}.
\newblock \emph{\bibinfo{journal}{Nature}} \textbf{\bibinfo{volume}{556}},
  \bibinfo{pages}{478--482} (\bibinfo{year}{2018}).

\bibitem{RiedingerNature2018}
\bibinfo{author}{Riedinger, R.} \emph{et~al.}
\newblock \bibinfo{title}{Remote quantum entanglement between two
  micromechanical oscillators}.
\newblock \emph{\bibinfo{journal}{Nature}} \textbf{\bibinfo{volume}{556}},
  \bibinfo{pages}{473--477} (\bibinfo{year}{2018}).

\bibitem{LeeScience2011}
\bibinfo{author}{Lee, K.~C.} \emph{et~al.}
\newblock \bibinfo{title}{Entangling macroscopic diamonds at room temperature}.
\newblock \emph{\bibinfo{journal}{Science}} \textbf{\bibinfo{volume}{334}},
  \bibinfo{pages}{1253--1256} (\bibinfo{year}{2011}).

\bibitem{holsteinprimakoff}
\bibinfo{author}{Holstein, T.} \& \bibinfo{author}{Primakoff, H.}
\newblock \bibinfo{title}{Field dependence of the intrinsic domain
  magnetization of a ferromagnet}.
\newblock \emph{\bibinfo{journal}{Phys. Rev.}} \textbf{\bibinfo{volume}{58}},
  \bibinfo{pages}{1098--1113} (\bibinfo{year}{1940}).

\bibitem{YanbeiChenPRA2009}
\bibinfo{author}{M\"uller-Ebhardt, H.} \emph{et~al.}
\newblock \bibinfo{title}{Quantum-state preparation and macroscopic
  entanglement in gravitational-wave detectors}.
\newblock \emph{\bibinfo{journal}{Phys. Rev. A}} \textbf{\bibinfo{volume}{80}},
  \bibinfo{pages}{043802} (\bibinfo{year}{2009}).

\bibitem{RossiPRL2019}
\bibinfo{author}{Rossi, M.}, \bibinfo{author}{Mason, D.},
  \bibinfo{author}{Chen, J.} \& \bibinfo{author}{Schliesser, A.}
\newblock \bibinfo{title}{Observing and verifying the quantum trajectory of a
  mechanical resonator}.
\newblock \emph{\bibinfo{journal}{Phys. Rev. Lett.}}
  \textbf{\bibinfo{volume}{123}}, \bibinfo{pages}{163601}
  (\bibinfo{year}{2019}).

\bibitem{WieczorekPRL2015}
\bibinfo{author}{Wieczorek, W.} \emph{et~al.}
\newblock \bibinfo{title}{Optimal state estimation for cavity optomechanical
  systems}.
\newblock \emph{\bibinfo{journal}{Phys. Rev. Lett.}}
  \textbf{\bibinfo{volume}{114}}, \bibinfo{pages}{223601}
  (\bibinfo{year}{2015}).

\bibitem{Wasilewski2010}
\bibinfo{author}{Wasilewski, W.} \emph{et~al.}
\newblock \bibinfo{title}{Quantum noise limited and entanglement-assisted
  magnetometry}.
\newblock \emph{\bibinfo{journal}{Phys. Rev. Lett.}}
  \textbf{\bibinfo{volume}{104}}, \bibinfo{pages}{133601}
  (\bibinfo{year}{2010}).

\bibitem{Balabas2010}
\bibinfo{author}{Balabas, M.~V.} \emph{et~al.}
\newblock \bibinfo{title}{High quality anti-relaxation coating material for
  alkali atom vapor cells}.
\newblock \emph{\bibinfo{journal}{Opt. Express}} \textbf{\bibinfo{volume}{18}},
  \bibinfo{pages}{5825--5830} (\bibinfo{year}{2010}).

\bibitem{Geremia2007}
\bibinfo{author}{Geremia, J.~M.}, \bibinfo{author}{Stockton, J.~K.} \&
  \bibinfo{author}{Mabuchi, H.}
\newblock \bibinfo{title}{Tensor polarizability and dispersive quantum
  measurement of multilevel atoms}.
\newblock \emph{\bibinfo{journal}{Phys. Rev. A}} \textbf{\bibinfo{volume}{73}},
  \bibinfo{pages}{042112} (\bibinfo{year}{2006}).

\bibitem{LammersThesis}
\bibinfo{author}{Lammers, J.}
\newblock \emph{\bibinfo{title}{State preparation and verification in
  continuously measured quantum systems}}.
\newblock Ph.D. thesis, \bibinfo{school}{Leibniz University Hannover}
  (\bibinfo{year}{2018}).

\bibitem{Julsgaard2004}
\bibinfo{author}{Julsgaard, B.}, \bibinfo{author}{Sherson, J.},
  \bibinfo{author}{S{\o}rensen, J.~L.} \& \bibinfo{author}{Polzik, E.~S.}
\newblock \bibinfo{title}{{Characterizing the spin state of an atomic ensemble
  using the magneto-optical resonance method}}.
\newblock \emph{\bibinfo{journal}{Journal of Optics B: Quantum and
  Semiclassical Optics}} \textbf{\bibinfo{volume}{6}}, \bibinfo{pages}{5}
  (\bibinfo{year}{2004}).

\bibitem{stokespar}
\bibinfo{author}{Agarwal, G.~S.} \& \bibinfo{author}{Chaturvedi, S.}
\newblock \bibinfo{title}{Scheme to measure quantum stokes parameters and their
  fluctuations and correlations}.
\newblock \emph{\bibinfo{journal}{Journal of Modern Optics}}
  \textbf{\bibinfo{volume}{50}}, \bibinfo{pages}{711--716}
  (\bibinfo{year}{2003}).

\bibitem{cifar}
\bibinfo{author}{{Thomas, R. A., \emph{et al.}}}
\newblock \bibinfo{title}{{Atomic Spin System Calibrations by Coherently
  Induced Faraday Rotation}}.
\newblock \bibinfo{note}{In preparation}.

\bibitem{OMITScience2010}
\bibinfo{author}{Weis, S.} \emph{et~al.}
\newblock \bibinfo{title}{Optomechanically induced transparency}.
\newblock \emph{\bibinfo{journal}{Science}} \textbf{\bibinfo{volume}{330}},
  \bibinfo{pages}{1520--1523} (\bibinfo{year}{2010}).

\bibitem{Jayich2008}
\bibinfo{author}{Jayich, A.~M.} \emph{et~al.}
\newblock \bibinfo{title}{{Dispersive optomechanics: A membrane inside a
  cavity}}.
\newblock \emph{\bibinfo{journal}{New Journal of Physics}}
  \textbf{\bibinfo{volume}{10}}, \bibinfo{pages}{095008}
  (\bibinfo{year}{2008}).

\bibitem{NielsenpPNAS2017}
\bibinfo{author}{Nielsen, W. H.~P.}, \bibinfo{author}{Tsaturyan, Y.},
  \bibinfo{author}{M{\o}ller, C.~B.}, \bibinfo{author}{Polzik, E.~S.} \&
  \bibinfo{author}{Schliesser, A.}
\newblock \bibinfo{title}{Multimode optomechanical system in the quantum
  regime}.
\newblock \emph{\bibinfo{journal}{Proceedings of the National Academy of
  Sciences}} \textbf{\bibinfo{volume}{114}}, \bibinfo{pages}{62--66}
  (\bibinfo{year}{2017}).

\bibitem{julsgaardthesis}
\bibinfo{author}{Julsgaard, B.}
\newblock \emph{\bibinfo{title}{Entanglement and Quantum Interactions with
  Macroscopic Gas Samples}}.
\newblock \bibinfo{type}{{PhD} dissertation}, \bibinfo{school}{University of
  Aarhus} (\bibinfo{year}{2003}).

\bibitem{WasilewskiOE2009}
\bibinfo{author}{Wasilewski, W.} \emph{et~al.}
\newblock \bibinfo{title}{Generation of two-mode squeezed and entangled light
  in a single temporal and spatial mode}.
\newblock \emph{\bibinfo{journal}{Opt. Express}} \textbf{\bibinfo{volume}{17}},
  \bibinfo{pages}{14444--14457} (\bibinfo{year}{2009}).

\bibitem{Carmichael1993}
\bibinfo{author}{Carmichael, H.}
\newblock \emph{\bibinfo{title}{An Open Systems Approach to Quantum Optics}}
  (\bibinfo{publisher}{Springer}, \bibinfo{address}{Berlin, Heidelberg},
  \bibinfo{year}{1993}).

\bibitem{Helstrom1976}
\bibinfo{author}{Helstrom, C.~W.}
\newblock \emph{\bibinfo{title}{Quantum detection and estimation theory}}
  (\bibinfo{publisher}{Academic Press}, \bibinfo{year}{1976}).

\bibitem{HaixingMiaoThesis}
\bibinfo{author}{Miao, H.}
\newblock \emph{\bibinfo{title}{Exploring Macroscopic Quantum Mechanics in
  Optomechanical Devices}}.
\newblock Ph.D. thesis, \bibinfo{school}{University of Western Australia}
  (\bibinfo{year}{2010}).

\bibitem{Broersen2006}
\bibinfo{author}{Broersen, P. M.~T.}
\newblock \emph{\bibinfo{title}{Automatic Autocorrelation and Spectral
  Analysis}} (\bibinfo{publisher}{Springer-Verlag}, \bibinfo{address}{Berlin,
  Heidelberg}, \bibinfo{year}{2006}).

\bibitem{wiener1964extrapolation}
\bibinfo{author}{Wiener, N.}
\newblock \emph{\bibinfo{title}{Extrapolation, Interpolation, and Smoothing of
  Stationary Time Series: With Engineering Applications}}.
\newblock Massachusetts Institute of Technology : Paperback series
  (\bibinfo{publisher}{M.I.T. Press}, \bibinfo{year}{1964}).

\bibitem{emcee}
\bibinfo{author}{Foreman-Mackey, D.}, \bibinfo{author}{Hogg, D.~W.},
  \bibinfo{author}{Lang, D.} \& \bibinfo{author}{Goodman, J.}
\newblock \bibinfo{title}{emcee: The {MCMC} hammer}.
\newblock \emph{\bibinfo{journal}{Publications of the Astronomical Society of
  the Pacific}} \textbf{\bibinfo{volume}{125}}, \bibinfo{pages}{306--312}
  (\bibinfo{year}{2013}).

\end{thebibliography}

\appendix
\clearpage

\clearpage
\onecolumngrid
\section*{Methods}

\section{Experimental setup}
\label{app:exp}
\subsection{Atomic spins}
\label{app:exp:spins}
The atomic spin oscillator is prepared in a \SI{50}{\celsius} warm ensemble of caesium atoms, confined in a spin anti-relaxation-coated microcell \citep{Balabas2010}
($\SI{300}{\micro\meter}\times\SI{300}{\micro\meter}$ cross-section and $\SI{10}{\milli\meter}$ in length). The natural linewidth, in the absence of light, is $\gamma_{\text{S}0,\textrm{dark}}=1/(\pi T_2)=\SI{450}{Hz}$, as measured by pulsed Magneto-Optical Resonance Signal (MORS)~\citep{Julsgaard2004}. 

The microcell is positioned in a magnetic shield equipped with coils producing a homogeneous magnetic bias field orthogonal to the probe direction, and a heater to keep the interior at the desired temperature, effectively determining the total atom number. The magnetic field direction sets the quantization axis, denoted as the \textit{x}-direction. The high thermal mass of the shield ensures a stable temperature throughout the experimental trials. The resonance frequency of the spin $|\omega_\text{S}|$, i.e., the Larmor frequency, is controlled by the magnitude of the magnetic field. 

The atoms travel through a Gaussian mode of the probe laser focused at the center of the microcell, with the beam waist ($w_0\approx \SI{80}{\micro\meter}$) optimized to maximize the filling factor without incurring extra optical losses. The laser frequency is blue-detuned by $\SI{3}{\giga\hertz}$ from the $F=4 \rightarrow F'=5$ D$_2$ transition. Even at this detuning the tensor interaction is non-negligible, which requires a careful choice of the input linear polarisation. The chosen polarisation is at the angle $\alpha \approx \SI{60\pm2}{\degree}$ with respect to the magnetic field such that the tensorial Stark shifts induced by the probe cancel the quadratic Zeeman splitting $\omega_\text{qzs}= \SI{400}{Hz}$, as described by the atomic polarisability tensor (see Methods~\ref{app:model:spins} for more details).  

The standard quantum Stokes variables ${\hat{S}_x,\hat{S}_y,\hat{S}_z,\hat{S}_0}$ -- representing the light electric field in terms of its linear, diagonal, and circular polarisation states~\citep{stokespar} and the total intensity -- are redefined as $\{\hat{S}_\parallel=\hat{S}_x\cos 2\alpha-\hat{S}_y\sin 2\alpha,\hat{S}_\perp = \hat{S}_x\sin 2\alpha+\hat{S}_y\cos 2\alpha,\hat{S}_z,\hat{S}_0\}$. 
When mapping the polarisation variables into quadrature variables, we choose the parallel component as the classical variable -- the local oscillator $\text{LO}_1$ with the photon flux $\langle \hat{S}_\parallel \rangle = \langle \hat{S}_0 \rangle = S_\parallel$, leaving ${\hat{S}_\perp, \hat{S}_z}$ as quantum variables. We define the light quadratures as $\hat{X}_\text{L}=\hat{S}_\text{z}/\sqrt{S_\parallel}$ and  $\hat{P}_\text{L}=-\hat{S}_\perp/\sqrt{S_\parallel}$. 
The photon flux determines the readout rate $\Gamma_\text{S}/2\pi\propto J_x S_\parallel$ and the power broadening decoherence rate $\gamma_\text{pb}/2\pi\propto S_\parallel$. In the experiment, the $\textrm{LO}_1$ power is $\sim \SI{350}{\micro\watt}$. The decoherence rate of the oscillator is also affected by the optical pumping process, represented by the contribution $\gamma_\text{op}$. The total bandwidth of the spin resonance in the absence of dynamical processes is $\gamma_{\textrm{S0}}=\gamma_\textrm{S0,dark}+\gamma_\text{pb}+\gamma_\text{op}$. For the conditions of the current experiment, the ratio of quantum backaction to thermal noise contributions is $\textrm{QBA/TH} = \QBATHS$.

As the atoms move in and out of the beam, the scattered photons couple to various atomic motional modes. The motion of the atoms is fast (flight-through time $\sim$\SI{1}{\micro\second}) and uncorrelated, leading to a motionally averaged coupling~\citep{Borregaard2016}. Phenomenologically, the long-lived correlations give rise to the mean spin mode -- the mode of interest -- and the short-time correlations to an uncorrelated spin contribution -- the broadband spin mode. In the regime of operation both optical responses are harmonic, with the susceptibility of short-time correlations following a low-$Q$ damped harmonic oscillator type, with resonance frequency $\Omega_\text{S}$ and linewidth $\gamma_\text{bb}/2\pi\sim\SI{1}{\mega\hertz}$ and coupling rate $\Gamma_\textrm{S,bb}$. 

\begin{figure}
    \centering
    \includegraphics[]{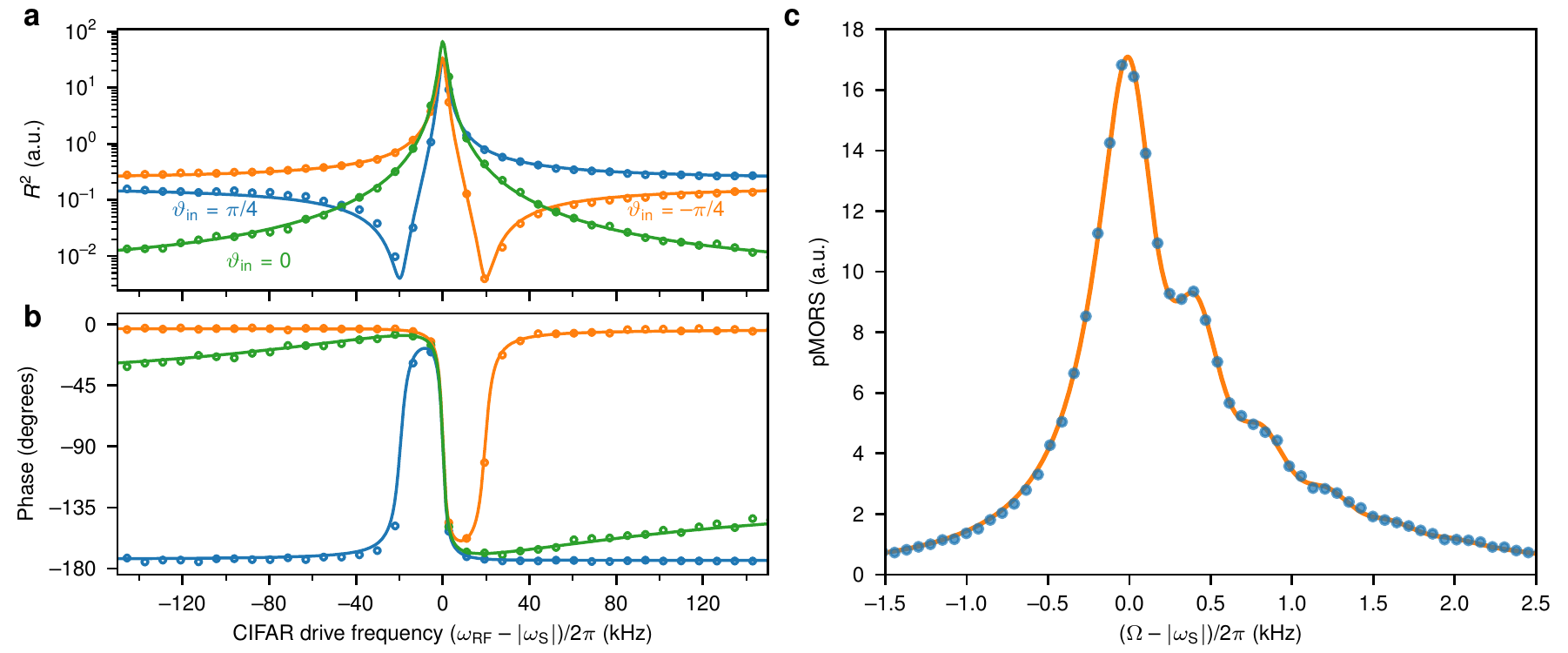}
    \caption{\textbf{Spin oscillator calibrations.} \textbf{a} and \textbf{b} show the square amplitude ($R^2$) and phase of the demodulated spin response to a coherently driven light field with different $\theta_\text{in}$. \textbf{c,} pulsed MORS spectrum for the distribution of atoms in the Zeeman levels corresponding to $p=\atomicPol$.}
    \label{fig:atomiccalibrations}
\end{figure}

We observe the response of the two spin modes to coherent drive tones $\hat{X}_\textrm{L,S}^\textrm{drive}$ in Figures~\ref{fig:atomiccalibrations}(a) and \ref{fig:atomiccalibrations}(b) for different input modulation types $\hat{X}_\textrm{L,S}^\textrm{drive}=\hat{X}_\textrm{L,S}^\textrm{in}\cos\vartheta_\textrm{in}+\hat{P}_\textrm{L,S}^\textrm{in}\sin\vartheta_\textrm{in}$, measured by Coherent Induced FAraday Rotation (CIFAR) \citep{cifar}, a calibration technique which is inspired by the OptoMechanically Induced Transparency (OMIT)~\citep{OMITScience2010}. In short, CIFAR references the phase-sensitive response of the spin to an oscillating input polarization at $\omega_{\mathrm{RF}}$; to the first order, the resulting interference between the drive and response, for $\Gamma_\textrm{S}/\gamma_S \gg 1$ and $\vartheta_\textrm{in}=\pm\pi/4$, gives rise to a dispersive feature in the detected field, with maximum destructive interference at $\pm\Gamma_\textrm{S}$ away from $|\omega_\textrm{S}|$. Under the assumption that both modes are uncorrelated, we fit these data to the input-output relations~(\eqref{XLSfinal}), allowing us to extract the readout rate $\Gamma_\textrm{S}$. The backaction on the broadband mode is negligible, i.e., $\Gamma_\textrm{S,bb}/\gamma_\textrm{S,bb} \ll 1$. For all noise spectra, we treat the broadband contribution as constant in the frequency range of interest, added incoherently with all other noise processes; effectively, it acts as added phase noise in the phase quadrature of light. The added spectral power at the resonance frequency due to extra spin noise, corrected for losses, is $\overline{S}_\textrm{S,bb}= \BBNoiseLevel\ \textrm{SN}$ units.

The spin oscillator is prepared by optically pumping the ensemble towards the $|F=4,m_F=4\rangle$ Zeeman sublevel. A repump laser is tuned to the  $F=3 \rightarrow F'=2$ hyperfine transition in the D$_2$ line and a pump laser to the  $F=4 \rightarrow F'=4$ hyperfine transition in the D$_1$ line, both circularly polarised. The pump laser directly couples to the coherences of interest, competing with the decoherence and depumping caused by the probe, adding $\gamma_\textrm{op}/2\pi\sim \SI{1}{kHz}$. The spin polarization $p=\atomicPol$ is characterised by pulsed MORS~\citep{Julsgaard2004}, with the spectrum shown in Figure~\ref{fig:atomiccalibrations}(c). Due to the dominant role of the probing and pumping lasers, the Zeeman population distribution does not follow the spin temperature model. The fitting model follows Eq.~(17) from Ref.~\citep{Julsgaard2004} for arbitrary Zeeman population distribution. The spin polarisation is determined by assuming that the population of $|F=4,m_F=-4\rangle$ is negligible, which is guaranteed by the presence of the resonant pump laser. The spin oscillator variables $\hat{X}_\text{S}=\hat{J}_z/\sqrt{\hbar J_x}$ and $\hat{P}_\text{S}=-\hat{J}_y/\sqrt{\hbar J_x}$ are defined according to the steady-state spin polarisation, which defines $J_x$. From the population distribution we calculate the variance of the spin components, which leads to the added spin thermal occupancy $n_\textrm{S}=\nS$, meaning that incoherent processes drive towards an equilibrium with $\text{Var}[\hat{X}_\textrm{S}]=\text{Var}[\hat{P}_\textrm{S}]=n_\textrm{S}+1/2 $.

The spin system can be operated in two regimes, which differ only by their effective masses. Optical pumping of the atoms to the highest energy state, i.e.~spins aligned parallel with the bias magnetic field, leads to an effective negative mass, whereas pumping to the lowest energy state, i.e.~spin aligned anti-parallel to the bias field, leads to an effective positive mass~\citep{Moller2017}. This choice defines the  sign of the atomic susceptibility $\chi_{\text{S}}$.

\subsection{Optomechanics}
\label{app:exp:mech}
The optomechanical system consists of a \SI{13}{\nano\meter} thick, highly stressed, phononically patterned silicon nitride membrane featuring a soft-clamped~\citep{softclampingTsaturyan2017}, localised out-of-plane vibrational mechanical mode with a cryogenic $Q$-factor of \num{0.65e9} and resonance frequency of \SI{1.37}{\mega\hertz}. This membrane is positioned close to the waist of a \SI{2.6}{\milli\meter} long optical cavity along its axis and with maximum spatial overlap between the cavity mode and the localized mechanical mode. The cavity consists of two mirrors with \SI{25}{\milli\meter} radius of curvature, and power transmissions of \SI{20}{ppm} and \SI{360}{ppm}, respectively. The entire optomechanical assembly is placed in a liquid helium flow cryostat, which is cooled to \SI{4.4}{\kelvin}.

The basis of the light-mechanics coupling is the radiation pressure force of light on the membrane whose out-of-plane motion causes a dispersive shift of the cavity resonance frequency~\cite{Jayich2008}. The placement of the membrane inside the cavity divides it into two sub-cavities, where the amount of light in each depends on the membrane position. This system can be formally treated as a canonical end-mirror optomechanical system with just a single intracavity optical field. In this formalism adjusting the lengths of each subcavity, periodically modulates the canonical cavity parameters of optical linewidth, $\kappa$, resonance frequency, overcoupling $\kappa_{\mathrm{in}} / \kappa$ (where $\kappa_\mathrm{in}$ is the coupling rate of the input mirror), as well as the optomechanical single-photon coupling rate $g_0$.

The sub-cavities can be independently and electronically fine-tuned so as to simultaneously realise a high cooperativity optomechanical system, as well as tunability, in order to set an appropriate cavity detuning with respect to the probe of the atomic spin system. The various canonical optomechanical parameters are characterised through several independent measurements and a full list of these system parameters can be seen in Table~\ref{table:parameters}.

The cavity linewidth is characterised first by measuring the optical amplitude quadrature beatnote of a phase-modulation sideband transmitted through the cavity. In a second method, a single carrier is scanned across the cavity resonance on a timescale comparable to the cavity response time and the resultant beating ringdown signal is observed.

The cavity detuning is determined by combining the characterisation of the cavity dither lock error signal and knowledge of the cavity linewidth. By locating the turning point of the dither error signal we translate our locked error signal amplitude into an absolute detuning.

\begin{figure}
    \centering
    \includegraphics[]{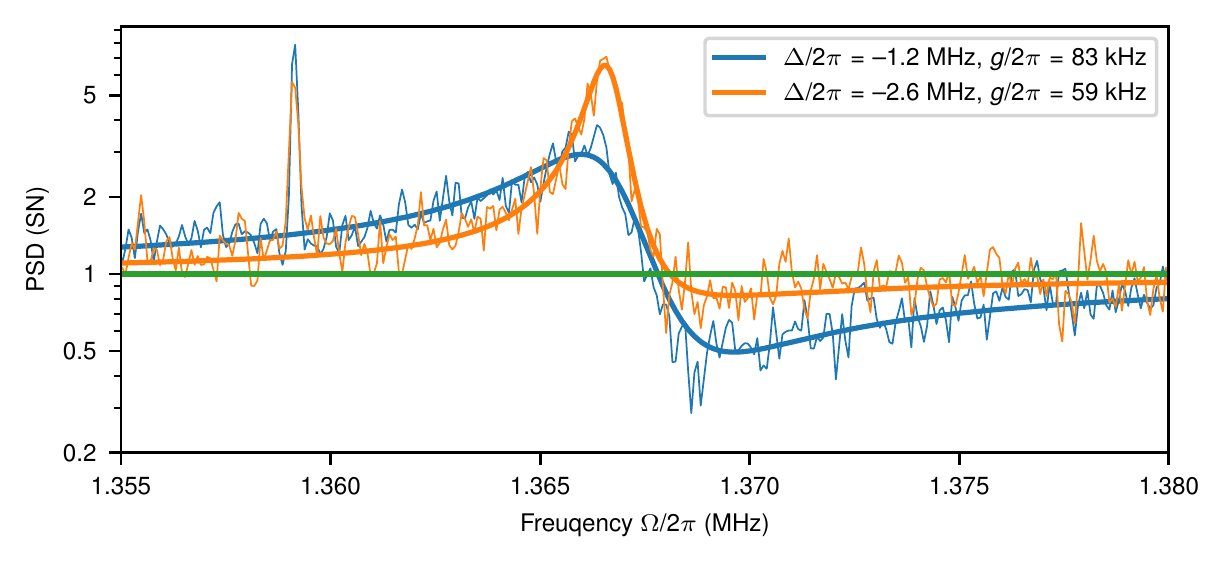}
    \caption{\textbf{Ponderomotive squeezing spectra for different cavity detunings.} From these spectra we infer $g$, $T$, optical losses in the detection path etc. See Methods~\ref{app:exp:mech} for details. Using other system parameters, measured independently, we obtain an effective bath temperature of $T=\SI[separate-uncertainty=true]{11.4\pm0.5}{\kelvin}$.
    }
    \label{fig:squeezing-tmep}
\end{figure}

The effective mechanical bath temperature and field-enhanced optomechanical coupling rate $g=g_0\vert\alpha\vert$, where $\alpha$ is intra-cavity field, can be obtained by fitting the full optomechanical model to the ponderomotive squeezing spectra, seen in Fig.~\ref{fig:squeezing-tmep}. These spectra result from pumping the cavity from the high-reflector port and detecting the optical amplitude quadrature in transmission through the highly overcoupled port~\cite{NielsenpPNAS2017}. We fit the model to these two spectra simultaneously, using  separately measured values for $\Delta$, $\kappa$, and $Q$. From this characterisation the detection efficiency, with and without \LO3, can similarly be inferred. We observed up to \SI{3}{\decibel} of ponderomotive squeezing. We note that while our system is not optimized for measuring maximum ponderomotive squeezing, nor operated in the optimum regime, we observe close to record amounts of squeezing for optomechanical systems.

A new feature of our optomechanical cavity compared to our previous work~\cite{Moller2017} is the full electronic control over the position of the membrane inside the standing wave of the cavity. Two piezos, each with an effective travel length of well over a half-wavelength at cryogenic conditions, allow us to scan the lengths of the two sub-cavities, so as to effectively position the membrane at any given intra-cavity position while keeping the cavity on resonance with the optical light field. By monitoring the cavity transmission as we scan the position of the membrane, we obtain knowledge about the position within the standing wave. We operate the optomechanical system at the point of highest total cavity linewidth, giving us the best overcoupling in reflection, as well as a high coupling rate $g_0$.

\subsection{Hybrid system matching \& homodyne detection}
\label{app:exp:hybrid}

The overall hybrid system consists of the cascaded optical readout of the spin and mechanical system by an itinerant light field, as outlined in the main text, see Fig.~\ref{fig:expsetup}.

After interacting with the spin, \LO1 is filtered off the quantum signal, which is orthogonally polarisation to \LO1. The quantum signal is spatially overlapped with \LO2 on a PBS, after which the two beams co-propagate, but have different polarisation. To remedy this, we use a $\lambda/2$ plate and a second PBS to reject most of the \LO2 beam and retain most of the quantum signal, incurring a small (percent-scale) loss of the quantum signal. 
This directly translates the polarization quadrature operators that interacted with the spin system into the amplitude and phase quadratures that are now coupled to the membrane-in-the-middle optomechanical system, in which radiation pressure of the \LO2 drives the mechanical oscillator.

The cascaded system, including the double pass nature of our atomic read out, makes our system susceptible to back-reflections from the optomechanical system to the spin system, since these reflections effectively amount to a self-driving force on the spins, leading to self-induced oscillations of the spin system. Therefore, the system necessitates the introduction of an optical isolator, leading to additional optical intersystem losses. Further, non-perfect rejection of \LO1 by the PBS separating \LO1 and the quantum signal leads to a part of \LO1 co-propagating with \LO2. These two LOs interfere, which effectively turns drifts in the \LO1-\LO2 phase $\varphi$ into changes of the total optical power sent to the mechanical system.

The optical output of the spin system is spatially mode matched to the optomechanical cavity by using \LO1 as a proxy. By rotating waveplates, the \LO1 probe is directed to the cavity and modematched to it. The degree of modematching is characterized by the amount of ponderomotive squeezing observed in the optical amplitude quadrature in reflection.

Phase fluctuations of the light reflected off the optomechanical cavity are measured  with homodyne detection as depicted in Fig.~\ref{fig:expsetup} of the main text.
The reflected beam is spatially overlapped with \LO3 on a polarising beamsplitter (PBS), and the \LO3 is mode matched to the optical signal. The \LO3 and \LO2 plus quantum fluctuations are now co-propagating, but in different polarisation channels. They are transmitted through a $\lambda/2$ waveplate, set to rotate the polarisations by \SI{45}{\degree}. The mixed polarisation components are then respectively transmitted and reflected off the second PBS. Neglecting interference, this splits both components equally into the two ports. The total set of PBS-$\lambda/2$-PBS thus acts as an effective 50:50 beamsplitter. The light fields are now in the correct polarisation channels to interfere for homodyne detection.

We perform differential detection, by measuring the photocurrents of a photodetector in each arm, and electronically subtracting the two currents. The slow component is fed back to a piezo, controlling the optical path in the \LO3 arm, thus determining the homodyne detection angle $\vartheta$. The optical powers are $\sim \SI{2}{\milli\watt}$ of \LO3 and $\sim \SI{9}{\micro\watt}$ of \LO2.

\begin{figure}
    \centering
    \includegraphics{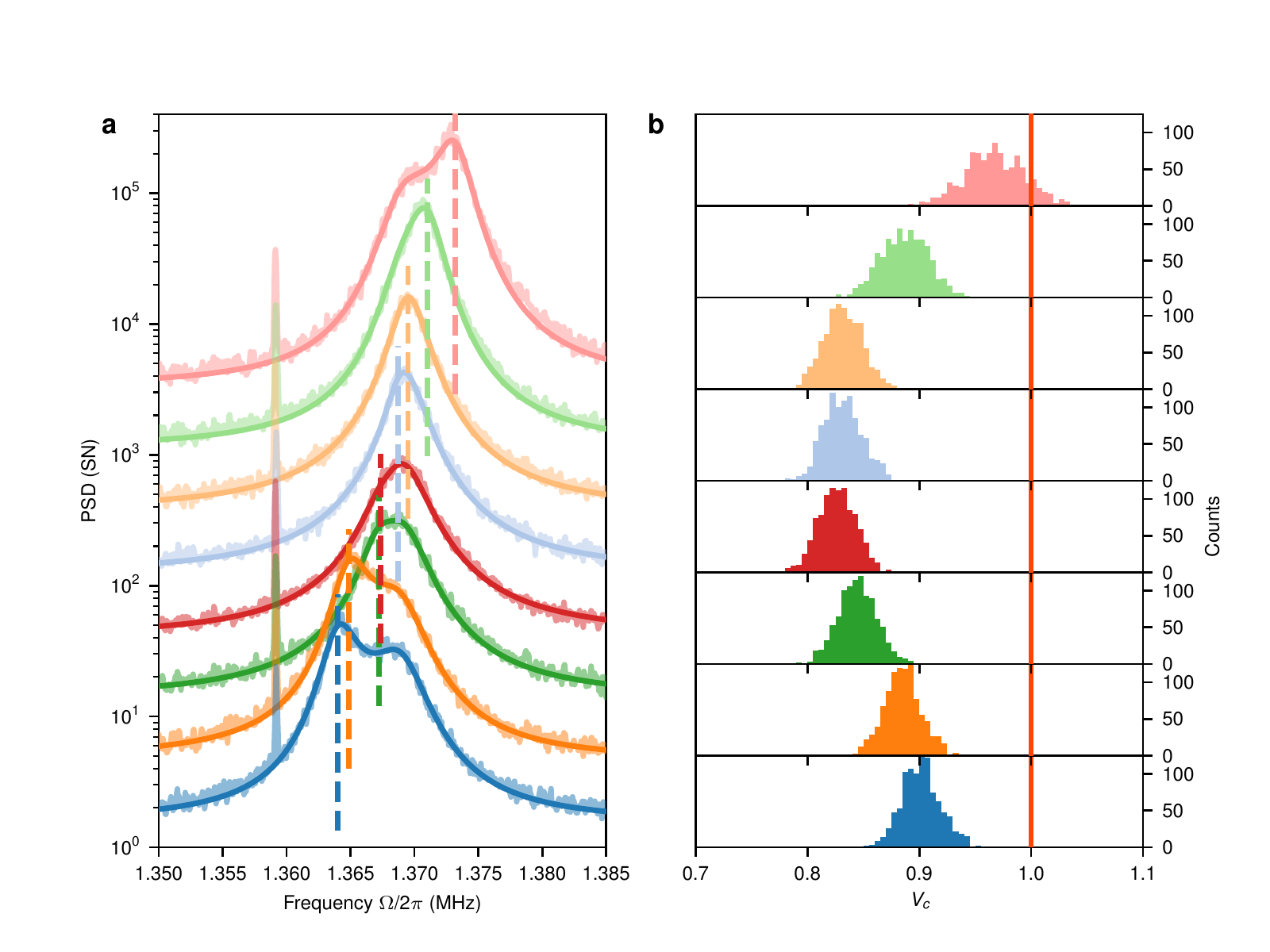}
    \caption{ \textbf{a,} Fit results for varied atomic frequencies $\omega_\mathrm{S}$ for all points shown in Fig.~\ref{fig:entanglement}. For clarity, subsequent lines are offset vertically by multiplying by a constant factor.  \textbf{b,} MCMC results for the conditional variance for all atomic detunings. Mean and standard deviation leads to the points in Fig.~\ref{fig:entanglement}b.}
    \label{fig:histograms}
\end{figure}

Experimental spectra are presented in the main text, Figs.~\ref{fig:spectra}~and~\ref{fig:entanglement}, as well as in Figs.~\ref{fig:WF-widerange}~and~\ref{fig:histograms}. In Fig.~\ref{fig:WF-widerange} we present a wider frequency range, thus showing features such as out-of-bandgap mechanical modes, mechanical modes of the mirror substrates, higher-order mechanical modes in the bandgap, etc. In Fig.~\ref{fig:histograms} we present experimental spectra plus model fits for all atomic detunings presented in Fig.~\ref{fig:entanglement}.

\section{Theoretical model}
\label{app:model}
In this section we will present the model used to fit the experimental data and to extract parameters necessary for the entanglement analysis and Wiener filtering. The latter also relies on signal and noise (cross-)correlation functions calculated from the (fitted) model.

For a function in the time domain $\hat{f}(t)$, we use the Fourier transform sign convention and property
\begin{align}\label{eq:Fourier-convention}
    \hat{f}(\Omega) =\mathcal{F}\{\hat{f}(t)\} = \int_{-\infty}^{\infty} \hat{f}(t)e^{i\Omega t}\ \mathrm{d}t, \qquad \mathcal{F}\left\{\dfrac{\mathrm{d}}{\mathrm{d}t}\hat{f}(t)\right\} = -i\Omega \hat{f}(\Omega).
\end{align}

For the localized optical cavity mode, we introduce the photon annihilation and creation operators obeying the commutation relation $[\hat a,\hat a^\dagger]=1$, and, in turn, the light amplitude and phase quadratures (suppressing the time/Fourier-frequency dependence for brevity)
\begin{align}
        \hat{X}_\textrm{L} = \dfrac{\hat{a}+\hat{a}^\dagger}{2} \qquad
        \hat{P}_\textrm{L} = \dfrac{\hat{a}-\hat{a}^\dagger}{2i},
\end{align}
which obey the same-time commutation relation $[\hat{X}_\textrm{L}(t) , \hat{P}_\textrm{L}(t) ]=i/2$. 

All traveling optical fields, including additional (vacuum) noise fields introduced by optical losses, are described by amplitude and phase quadratures
\begin{align}
        \hat{X}_{\textrm{L}}^{\text{in(out)}} = \dfrac{\hat{a}_{\text{in(out)}}^{\phantom{\dagger}}+\hat{a}_{\text{in(out)}}^\dagger}{2} \qquad
        \hat{P}_{\textrm{L}}^{\text{in(out)}} = \dfrac{\hat{a}_{\text{in(out)}}^{\phantom{\dagger}}-\hat{a}_{\text{in(out)}}^\dagger}{2i},
\end{align}
defined in terms of the quantum amplitudes
\begin{equation}\label{eq:a-inout-F-decomp}
\hat{a}_{\text{in(out)}}(t)=\frac{1}{2\pi}\int_{-\infty}^{\infty}\mathrm{d}\Omega\ e^{-i\Omega t}\hat{a}_{\text{in(out)}}(\Omega)\qquad \hat{a}_{\text{in(out)}}^{\dagger}(t)=\frac{1}{2\pi}\int_{-\infty}^{\infty}\mathrm{d}\Omega\ e^{+i\Omega t}\hat{a}_{\text{in(out)}}^{\dagger}(\Omega)
\end{equation}
where $\hat{a}_{\text{in(out)}}$ is the field in a rotating frame with respect to the relevant optical carrier frequency $\omega_\textrm{laser}$, so that $\hat{a}_{\text{in(out)}}(\Omega)$ represents the field at absolute frequency $\Omega+\omega_\textrm{laser}$. This expression is valid for Fourier frequencies close to the optical carrier, $|\Omega| \ll \omega_\textrm{laser}$. According to the above considerations the Fourier transforms of the
rotating-frame operators $\hat{a}_{\text{in(out)}}(t)$ and $\hat{a}_{\text{in(out)}}^{\dagger}(t)$~[Eqs.~(\ref{eq:a-inout-F-decomp})],
using the convention in Eq.~(\ref{eq:Fourier-convention}), are
\begin{equation}
\mathcal{F}\{\hat{a}_{\text{in(out)}}(t)\}=\hat{a}_{\text{in(out)}}(\Omega),\quad\mathcal{F}\{\hat{a}_{\text{in(out)}}^{\dagger}(t)\}=\hat{a}_{\text{in(out)}}^{\dagger}(-\Omega).
\end{equation}
The non-vanishing commutation relations of the travelling field operators are $[\hat{X}_{\text{L}}^{\text{in(out)}}(t),\hat{P}_{\text{L}}^{\text{in(out)}}(t')]=(i/2)\delta(t-t')$. Accordingly, the symmetrised power spectral densities of the incoming vacuum light fields are
\begin{subequations}\label{eq:psdlight}
\begin{align}
    \overline{S}_{X_\textrm{L} X_\textrm{L}}(\Omega)\delta(\Omega-\Omega') = \frac{1}{2}\langle \hat{X}_{\text{L},j}^{\text{in}\dagger}(\Omega)\hat{X}_{\text{L},j}^{\text{in}}(\Omega')+\hat{X}_{\text{L},j}^{\text{in}}(\Omega')\hat{X}_{\text{L},j}^{\text{in}\dagger}(\Omega)\rangle= \frac{1}{4}\delta(\Omega-\Omega')\\  \overline{S}_{P_\textrm{L} P_\textrm{L}}(\Omega)\delta(\Omega-\Omega') = \frac{1}{2}\langle \hat{P}_{\text{L},j}^{\text{in}\dagger}(\Omega)\hat{P}_{\text{L},j}^{\text{in}}(\Omega')+\hat{P}_{\text{L},j}^{\text{in}}(\Omega')\hat{P}_{\text{L},j}^{\text{in}\dagger}(\Omega)\rangle= \frac{1}{4}\delta(\Omega-\Omega').
\end{align}
\end{subequations}

\begin{figure}
    \centering
    \includegraphics{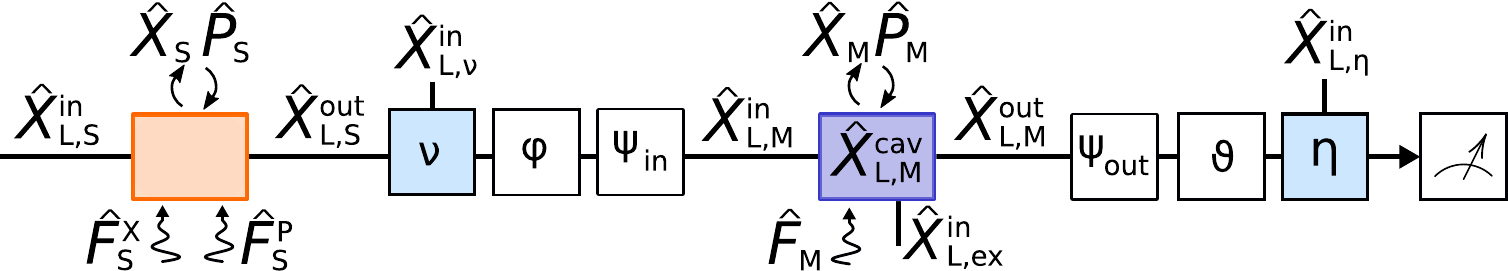}
    \caption{\textrm{Diagramatic representation.} Various optical fields, operators, thermal bath forces and rotations acting in the hybrid system, from input to detection. Spin (orange box) and mechanical system (blue box) along with driving optical and thermal forces. Light blue boxes represent beam-splitter-like losses. White boxes represent the various rotations applied to the optical fields.} 
    \label{fig:variables}
\end{figure}

For the mechanical (M) and spin (S) oscillators, we follow the commutation relation $[\hat X_j,\hat P_j]=i$ for $ (j=\text{M},\text{S})$; the effect of the thermal reservoirs $\hat{F}_j$ with mean thermal occupancy $n_j$ is captured by the symmetrised correlation functions  
\begin{subequations}
\label{eq:psdthermalforces}
\begin{align}
   \overline{S}_{F_S^X F_S^X}(\Omega)\delta(\Omega-\Omega') &\equiv \dfrac{1}{2}\langle \hat{F}_\textrm{S}^{\textrm{X},\dagger}(\Omega)\hat{F}^\textrm{X}_\textrm{S}(\Omega') + \hat{F}^\textrm{X}_\textrm{S}(\Omega')\hat{F}_\textrm{S}^{\textrm{X},\dagger}(\Omega) \rangle = \gamma_\text{S0}(n_\text{S}+1/2)\delta(\Omega-\Omega') \\
   \overline{S}_{F_S^P F_S^P}(\Omega)\delta(\Omega-\Omega') &\equiv \dfrac{1}{2}\langle \hat{F}_\textrm{S}^{\textrm{P},\dagger}(\Omega)\hat{F}^\textrm{P}_\textrm{S}(\Omega') + \hat{F}^\textrm{P}_\textrm{S}(\Omega')\hat{F}_\textrm{S}^{\textrm{P},\dagger}(\Omega) \rangle = \gamma_\text{S0}(n_\text{S}+1/2)\delta(\Omega-\Omega') \\
    \overline{S}_{F_M F_M}(\Omega)\delta(\Omega-\Omega') &\equiv \dfrac{1}{2}\langle \hat{F}_\textrm{M}^{\dagger}(\Omega)\hat{F}_\textrm{M}(\Omega') + \hat{F}_\textrm{M}(\Omega')\hat{F}_\textrm{M}^{\dagger}(\Omega) \rangle = 2\gamma_\text{M0}(n_\text{M}+1/2)\delta(\Omega-\Omega').
\end{align}
\end{subequations}
The diagrammatic representation of the fields and operations under considerations is presented in Fig.~\ref{fig:variables}. 

\subsection{Atomic ensemble}
\label{app:model:spins}

The atomic ensemble interacts dispersively with the light, leading to a mutual rotation of the light and spin variables according to the atomic polarizability tensor~\cite{julsgaardthesis}
\begin{align}
    \hat{H}_\text{S}/\hbar = -\omega_\text{S} \hat{J}_x + g_\textrm{S} \bigg[ a_0 \hat{S}_0 \hat{J}_0 + a_1 \hat{S}_z\hat{J}_z + 2a_2 \left[ \hat{S}_0\hat{J}_z^2 - \hat{S}_{x}(\hat{J}_{x}^2-\hat{J}_{y}^2) - \hat{S}_{y}(\hat{J}_x\hat{J}_y + \hat{J}_y \hat{J}_x)\right]\bigg],
\end{align}

with $a_0$, $a_1$, and $a_2$ as the relative weights of the scalar, vector and tensor contributions~\citep{julsgaardthesis}, which can be tuned by the detuning of the laser with respect to the atomic resonance and $g_\textrm{S}$ is the coupling rate. 
We work detuned $\SI{3}{\giga\hertz}$ to the blue from the $F=4 \rightarrow F'=5$ D$_2$ transition. 

In the limit of high spin polarization in the $F=4$ hyperfine manifold and for a strong linearly polarized local oscillator polarized at an angle $\alpha$ to the quantization axis, the Hamiltonian can be simplified to~\cite{WasilewskiOE2009} 
\begin{align}
    \hat{H}_\text{S}/\hbar &=\dfrac{ \omega_\text{S} }{2}(\hat{X}_\text{S}^2+\hat{P}_\text{S}^2) - 2\sqrt{\Gamma_\text{S}}\left(\hat{X}_\text{S} \hat{X}_\text{L} + \zeta_{\text{S}} \hat{P}_\text{S} \hat{P}_\text{L} \right),
\end{align}
where $\Gamma_\text{S} = g_\textrm{S}a_1 \sqrt{S_\parallel J_x}$ is the spin oscillator readout rate and $\zeta_{\text{S}} = -14  \frac{a_2}{a_1} \cos2\alpha$ is the tensor correction factor, which for our choice of polarization angle $\alpha$ has a value of $\sim \zetaS$. We have omitted constant energy terms, as they do not affect the dynamics of the spin variables of interest.  The canonical light variables are $\{\hat{X}_\text{L}=\hat{S}_\text{z}/\sqrt{S_\parallel},\hat{P}_\text{L}=-\hat{S}_\perp/\sqrt{S_\parallel}\}$. In our experimental regime, as $\zeta > 0$, the spin-light interactions deviates from the QND interaction, introducing extra correlation terms and allowing for dynamical cooling of the spin ensemble, effectively changing the decay rate and bath occupation. 

The dynamics follows from the Heisenberg-Langevin equations, which, in the steady state and in the frequency space, are
\begin{align}
   \left(\begin{array}{cc}
        \gamma_\text{S0}/2+\zeta_{\text{S}}\Gamma_\text{S}-i\Omega & -\omega_\text{S} \\
        \omega_\text{S} & \gamma_\text{S0}/2+\zeta_{\text{S}}\Gamma_\text{S}-i\Omega
    \end{array}\right)
    \left(\begin{array}{c}
        \hat{X}_\text{S} \\
        \hat{P}_\text{S}
    \end{array}\right) &=  2\sqrt{\Gamma_\text{S}}
    \left(\begin{array}{cc}
        0 & -\zeta_{\text{S}} \\
        1 & 0
    \end{array}\right)
    \left(
    \begin{array}{c}
    \hat{X}_\text{L,S}^\text{in} \\
    \hat{P}_\text{L,S}^\text{in}
    \end{array}
    \right) + \left(
    \begin{array}{c}
    \hat{F}_\text{S}^\text{X} \\
    \hat{F}_\text{S}^\text{P}
    \end{array}
    \right), \\
    \left(
    \begin{array}{c}
    \hat{X}_\text{L,S}^\text{out} \\
    \hat{P}_\text{L,S}^\text{out}
    \end{array}
    \right) &= \left(
    \begin{array}{c}
    \hat{X}_\text{L,S}^\text{in} \\
    \hat{P}_\text{L,S}^\text{in}
    \end{array}
    \right) + \sqrt{\Gamma_\text{S}}
    \left(\begin{array}{cc}
        0 & -\zeta_{\text{S}} \\
        1 & 0
    \end{array}\right)
   \left(\begin{array}{c}
        \hat{X}_\text{S} \\
        \hat{P}_\text{S}
    \end{array}\right),
\end{align}
for $2\zeta_{\text{S}}\Gamma_\text{S}$ as the tensor (dynamical) broadening, and ${\hat{F}^\text{X}_\text{S},\hat{F}^\text{P}_\text{S}}$ as the effective force acting on the spins via the thermal bath. We proceed defining the shorthand matrix notation
\begin{gather}
\mathbf{Z}=\left(\begin{array}{cc}
0 & -\zeta_{\text{S}}\\
1 & 0
\end{array}\right),\quad\mathbf{L}=\left(\begin{array}{cc}
\gamma_{\text{S0}}/2+\zeta_{\text{S}}\Gamma_{\text{S}}-i\Omega & -\omega_{\text{S}}\\
\omega_{\text{S}} & \gamma_{\text{S0}}/2+\zeta_{\text{S}}\Gamma_{\text{S}}-i\Omega
\end{array}\right)^{-1},\nonumber\\
\hat{\bm{X}}_{\text{L,S}}^\textrm{in(out)}=\left(\begin{array}{c}
\hat{X}_{\text{L,S}}^\textrm{in(out)}\\
\hat{P}_{\text{L,S}}^\textrm{in(out)}
\end{array}\right),\quad\hat{\bm{X}}_{\text{S}}=\left(\begin{array}{c}
\hat{X}_{\text{S}}\\
\hat{P}_{\text{S}}
\end{array}\right),\quad \hat{\bm{F}}_\text{S}= \left(
    \begin{array}{c}
    \hat{F}_\text{S}^\text{X} \\
    \hat{F}_\text{S}^\text{P}
    \end{array}
    \right),
\end{gather}
and solve the equations for the atomic and light variables
\begin{align}
\hat{\bm{X}}_\text{S}&=2\sqrt{\Gamma_\text{S}}\mathbf{L}\mathbf{Z} \hat{\bm{X}}_\text{L,S}^\text{in}+\mathbf{L}\hat{\bm{F}}_\text{S} \label{eq:XSfinal}\\
\hat{\bm{X}}_\text{L,S}^\text{out}&=\hat{\bm{X}}_\text{L,S}^\text{in} + \sqrt{\Gamma_{\text{S}}}\mathbf{Z}\hat{\bm{X}}_\text{S}=(\mathbf{1}_2+2\Gamma_\text{S}\mathbf{Z} \mathbf{L}\mathbf{Z})\hat{\bm{X}}_\text{L,S}^\text{in}+\sqrt{\Gamma_\text{S}}\mathbf{Z}\mathbf{L}\hat{\bm{F}}_\text{S}\label{eq:XLSfinal},
\end{align}
where $\mathbf{1}_2$ is the $2\times 2$ identity matrix. 

In the main text we consider simpler, approximate versions of \eqref{XSfinal,XLSfinal} valid in the limit $|\omega_{\textrm{S}}|\gg \gamma_\textrm{S},|\Omega-|\omega_{\textrm{S}}||$. In this limit, the effective thermal forces $\hat{F}_\text{S}^\text{X}$ and $\hat{F}_\text{S}^\text{P}$
can be combined into the single thermal force term $\hat{F}_\text{S}\approx i\hat{F}_\text{S}^\text{X}+\hat{F}_\text{S}^\text{P}$.
In this limit, the evolution equation for $\hat{X}_\text{S}$ in terms of the susceptibility $\chi_{\text{S}}(\Omega)$ arises from Eq.~(\ref{eq:XSfinal}) (setting $\omega_{\text{S}0}\equiv\omega_{\text{S}}$),
\begin{equation}
    \hat{X}_{\text{S}}=\chi_{\text{S}}\left[\hat{F}_{\text{S}}+2\sqrt{\Gamma_{\text{S}}}\left(\begin{array}{cc}
        1 \\
        -i\zeta_\text{S}  
    \end{array}\right)^{\intercal}\hat{\bm{X}}_{\text{L,S}}^{\text{in}}\right]=\chi_{\text{S}}[\hat{F}_{\text{S}}+2\sqrt{\Gamma_{\text{S}}}(\hat{X}_{\text{L,S}}^{\text{in}}- i\zeta_{\text{S}}\hat{P}_{\text{L,S}}^{\text{in}})],
\end{equation}
as presented in the main text.  Noting that $\hat{P}_{\text{S}} \approx - \textrm{sign}(\omega_\textrm{S0})i \hat{X}_{\text{S}}$, the simpler input-output relation discussed in the main text, 
\begin{equation}
    \bm{X}_{\text{L,S}}^{\text{out}}=\bm{X}_{\text{L,S}}^{\text{in}}+\sqrt{\Gamma_{\text{S}}}\left(\begin{array}{c}
-i\zeta_{\text{S}}\\
1
\end{array}\right)\hat{X}_{\text{S}},\label{eq:spin-IO-simp-appendix}
\end{equation}
follows from Eq.~(\ref{eq:XLSfinal}).

The CIFAR modeling (see Methods~\ref{app:exp:spins}) is based on \eqref{XSfinal,XLSfinal}, with the broadband response added as another atomic mode in the following manner
\begin{align}
    \hat{\mathbf{X}}_\text{L,S}^\text{out}&=\hat{\mathbf{X}}_\text{L,S}^\text{in} + \sqrt{\Gamma_{\text{S}}}\mathbf{Z}\hat{\mathbf{X}}_\text{S}+ \sqrt{\Gamma_{\text{S,bb}}}\mathbf{Z}\hat{\mathbf{X}}_\text{S,bb} \\
    \hat{\mathbf{X}}_\text{S,bb}&=2\sqrt{\Gamma_\text{bb}}\mathbf{L}_\textrm{bb}\mathbf{Z} \hat{\mathbf{X}}_\text{L,S}^\text{in},
\end{align}
for $\mathbf{L}_\textrm{bb}$ as $\mathbf{L}$ with $\gamma_\textrm{S0}\rightarrow\gamma_\textrm{bb}$, $\Gamma_\textrm{S}\rightarrow\Gamma_\textrm{S,bb}$ and $\Gamma_{\text{S,bb}}$ as the broadband response readout rate. Incoherent thermal contributions were disregarded as the input field is modulated with large amplitude. The input field $\hat{\mathbf{X}}_\text{L,S}^\text{in}$ quadratures is rotated according to $\hat{\mathbf{X}}_\text{L,S}^\text{drive}=O_{\vartheta_\textrm{in}}\hat{\mathbf{X}}_\text{L,S}^\text{in}$, in which $\bm{O}_\alpha$
\begin{align}\label{eq:rotationmatrix}
{\bm O}_{\alpha}=\left(\begin{array}{cc}
         \cos\alpha & -\sin\alpha \\ 
         \sin\alpha & \cos\alpha
        \end{array}\right)
\end{align}
is a rotation matrix. The result of the CIFAR modelling for various $\vartheta_\textrm{in}$ is presented in Fig.~\ref{fig:atomiccalibrations}. \ref{fig:atomiccalibrations}(a) shows the amplitude squared $R^2$ of the detected field; \ref{fig:atomiccalibrations}(b) presents the phase of detected field in respect to the drive.

\subsection{Optomechanics}
\label{app:model:mech}

We start with the standard linearized optomechanical interaction between a mechanical degree of freedom with frequency $\omega_\text{M}$ and the intracavity field
 \begin{align}
 \hat{H}_\text{M}/\hbar &=\frac{\omega_\text{M}}{2}\left(\hat X_\text{M}^2+\hat P_\text{M}^2\right) - \Delta\left({\hat{X}_\text{L,M}^\text{cav}}{}^2+{\hat{P}_\text{L,M}^{\text{cav}}}{}^2\right)  - 4g \left(\hat X_\text{L,M}^\text{cav} \cos\psi_\text{in}+\hat P_\text{L,M}^\text{cav} \sin\psi_\text{in}\right)\hat X_\text{M},
 \end{align}
 where $\Delta=\omega_\text{L}-\omega_\text{c}$ is the detuning of the laser with respect to the cavity resonance $\omega_\text{c}$ and $g$ is the light-enhanced optomechanical coupling rate. The cavity linewidth $\kappa$ has contributions from the the in-and-out-coupling mirror ($\kappa_\mathrm{in}$) -- we probe the cavity in reflection -- and the highly-reflective (HR) back mirror ($\kappa_\mathrm{ex}^\text{HR}$) as well as from intracavity losses ($\kappa_\mathrm{ex}^\text{loss}$), such that $\kappa=\kappa_\mathrm{in}+\kappa_\text{ex}$, with $\kappa_\textrm{ex}=\kappa_\mathrm{ex}^\text{HR}+\kappa_\mathrm{ex}^\text{loss}$ where the subscript $\mathrm{ex}$ signifies any extra loss mechanism. Losses due to the HR mirror and due to intracavity scattering are mathematically equivalent. Finally, $\psi_\text{in}=\arctan(2\Delta/\kappa)$ denotes the phase of the intracavity field relative to input field. 
 
 The time evolution of the optical and mechanical variables, including decay and fluctuations, is given by the Heisenberg-Langevin equations. In the frequency domain, and in the steady-state regime, the equations of motion are
\begin{align}\label{eq:omevomatrix}
\begin{pmatrix}
\kappa/2-i\Omega & \Delta & 2g\sin\psi_\text{in}\\
-\Delta & \kappa/2-i\Omega & -2g\cos\psi_\text{in}\\
-4g\cos\psi_\text{in} & -4g\sin\psi_\text{in} & {\chi_{\text{M00}}^{-1}}
\end{pmatrix}
\begin{pmatrix}
\hat{X}_{\text{L,M}}^{\text{cav}}\\
\hat{P}_{\text{L,M}}^{\text{cav}}\\
\hat{X}_{\text{M}}
\end{pmatrix}
=
\begin{pmatrix}
\sqrt{\kappa_\mathrm{in}}\hat{X}_{\text{L,M}}^{\text{in}}+\sqrt{\kappa_\mathrm{ex}}\hat{X}_{\text{L,M}}^{\text{ex}}\\
\sqrt{\kappa_\mathrm{in}}\hat{P}_{\text{L,M}}^{\text{in}}+\sqrt{\kappa_\mathrm{ex}}\hat{P}_{\text{L,M}}^{\text{ex}}\\
\hat{F}_{\mathrm{M}}
\end{pmatrix},
\end{align}
in which $\chi_\text{M00}^{-1}\equiv(\omega_\text{M0}^2-\Omega^2-i\Omega\gamma_{\text{M0}})/\omega_{\text{M}0}$ (the subscript denotes that this susceptibility excludes both dynamical broadening and optical spring effects) and $\hat X_\text{L,M}^{\text{in}} $ ($\hat X_{\text{L,M}}^\text{ex}$) is the input quantum field leaking in via the port  `in' (`ex'). The port `in' corresponds to the main in/outcoupler, while mathematically port `ex' corresponds to both the HR mirror and intra-cavity loss, which act in the same way since no light is present at the input of HR. The dynamics of the membrane momentum are calculated from the relation $-i\Omega \hat{X}_\text{M}=\omega_\text{M0}\hat{P}_\text{M}$. The natural linewidth of the mechanical mode is $\gamma_\text{M0}$, and the mean occupation due to the thermal reservoir at temperature $T$ is $n_\text{M0}=\hbar \omega_\text{M0}/k_B T$.

We are interested both in the effect of the mechanical mode on the light variables and in the dynamics of the oscillator itself. By defining the matrices
\begin{align}
{\mathbf A}=\left(\begin{array}{cc} 
        \kappa/2-i\Omega & \Delta \\
        -\Delta & \kappa/2-i\Omega
    \end{array}\right), \quad
{\mathbf B}=\left(\begin{array}{c}
        0 \\
        -2g
    \end{array}\right),\quad
{\mathbf C}=\left(\begin{array}{cc}
        -4g & 0
    \end{array}\right), \quad
\hat{\bm{X}}_{\text{L,M}}^{j}=\left(\begin{array}{c}
\hat{X}_{\text{L,M}}^{j}\\
\hat{P}_{\text{L,M}}^{j}
\end{array}\right),
\end{align}
${\bm O}_\psi$ as the input-intracavity field phase rotation (see Eq.~(\ref{eq:rotationmatrix})) and the index ${j}\in\{\textrm{cav},\textrm{in},\textrm{ex}\}$ for optical fields, we write Eq.~(\ref{eq:omevomatrix}) as system of matrix equations. Noting that the cavity response matrix $\mathbf{A}$ is invariant under quadrature rotations, ${\bm O}_\psi \mathbf{A} {\bm O}^\intercal_\psi=\mathbf{A}$, we find the intracavity field and the mechanical variable as a function of the input fluctuations and thermal bath
\begin{align}  
\hat{{\bm X}}_{\text{L,M}}^\text{cav}&=\mathbf{A}^{-1} \left(\sqrt{\kappa_\text{in}}\hat{{\bm X}}_{\text{L,M}}^{\text{in}}+\sqrt{\kappa_\text{ex}}\hat{{\bm X}}_{\text{L,M}}^\text{ex}\right) - \mathbf{A}^{-1} \mathbf{O}_{\psi_\text{in}} \mathbf{B}\hat{X}_{\text{M}}, \label{eq:XLMcav-readout}\\
\hat{X}_{\text{M}}&=\chi_{\text{M}}\left[-\mathbf{C}\mathbf{A}^{-1}\mathbf{O}_{\psi_\text{in}}^{\intercal}\left(\sqrt{\kappa_{\text{in}}}\hat{{\bm{X}}}_{\text{L,M}}^{\text{in}}+\sqrt{\kappa_{\text{ex}}}\hat{{\bm{X}}}_{\text{L,M}}^{\text{ex}}\right)+\hat{F}_{\text{M}}\right]\label{eq:Xm-chiEff},
\end{align}
in which $\chi_{\text{M}}=(\chi_{\mathrm{M}00}^{-1}-\mathbf{CA}^{-1}\mathbf{B})^{-1}$ is the effective mechanical susceptibility in the presence of optomechanical coupling. Substituting Eq.~(\ref{eq:Xm-chiEff}) in Eq.~(\ref{eq:XLMcav-readout}) solves the system for the cavity field 
\begin{align}  
\hat{{\bm X}}_{\text{L,M}}^\text{cav}&={\mathbf O}_{\psi_\text{in}}{\mathbf Y}^{-1}{\mathbf O}_{\psi_\text{in}}^{\intercal}\left(\sqrt{\kappa_\text{in}}\hat{{\bm X}}_{\text{L,M}}^{\text{in}}+\sqrt{\kappa_\text{ex}}\hat{{\bm X}}_{\text{L,M}}^\text{ex}\right)-\mathbf{O}_{\psi_\text{in}}\mathbf{Y}^{-1}\mathbf{B} \chi_{\mathrm{M}00} \hat{F}_\text{M},\label{eq:XLcavfinal}
\end{align}
where $\mathbf{Y}=\mathbf{A} - \mathbf{B} \chi_{\mathrm{M}00} \mathbf{C}$ is the effective cavity response matrix in the presence of optomechanical coupling. This quantity can also be used to express the mechanical response~(\ref{eq:Xm-chiEff}) as
\begin{align}  
\hat{X}_{\text{M}}&=-\chi_{\text{M00}}\mathbf{CY}^{-1}\bm{\mathbf{O}}_{\psi_{\text{in}}}^{\intercal}\left(\sqrt{\kappa_{\text{in}}}\hat{{\bm{X}}}_{\text{L,M}}^{\text{in}}+\sqrt{\kappa_{\text{ex}}}\hat{{\bm{X}}}_{\text{L,M}}^{\text{ex}}\right)+\chi_{\text{M}}\hat{F}_{\text{M}}.\label{eq:Xm-chiM00}
\end{align}

Finally, we detect the reflected field off port $1$ in a homodyne measurement. The phase of the outgoing classical carrier field with respect to the cavity field is given by $\psi_\text{out}=\arctan(2\Delta/(\kappa_\text{in}-\kappa_\text{ex}))$. Overall, the total phase shift with respect to the input field is $\psi_\text{out}+\psi_\text{in}$. The cavity input-output relations, taking account for the acquired phase shift with respect to the input, from Eq.~(\ref{eq:XLcavfinal}), is
\begin{align}
\hat{\bm{X}}_{\mathrm{L,M}}^\textrm{out}&=\mathbf{O}_{\psi_\text{in}+\psi_\text{out}}^\intercal(-\hat{\bm{X}}_{\mathrm{L,M}}^{\textrm{in}}+\sqrt{{\kappa_\text{in}}}\hat{\bm{X}}_{\mathrm{L,M}}^\text{cav})\nonumber\\ &=\mathbf{O}_{\psi_\text{out}}^\intercal(\kappa_{\text{in}}\mathbf{Y}^{-1}-\mathbf{1}_2)\mathbf{O}_{\psi_\text{in}}^{\intercal}\hat{\bm{X}}_{\mathrm{L,M}}^{\textrm{in}}+\sqrt{\kappa_\text{in}{\kappa_\text{ex}}}\mathbf{O}_{\psi_\text{out}}^\intercal\mathbf{Y}^{-1}\mathbf{O}_{\psi_\text{in}}^{\intercal}\hat{\bm{X}}_{\mathrm{L,M}}^{\textrm{ex}}-\sqrt{\kappa_{\text{in}}}\mathbf{O}_{\psi_\text{out}}^\intercal\mathbf{Y}^{-1}\mathbf{B} \chi_\textrm{M00}\hat{F}_\text{M},\label{eq:XLMout-sol}
\end{align}
where in the second line we have substituted the solution for the intracavity field~(\ref{eq:XLcavfinal}).

Above we have developed the exact Fourier-domain solution to a (linearized) cavity-optomechanical system, in particular the mechanical response~(\ref{eq:Xm-chiEff}) and the optomechanical input-output relation~(\ref{eq:XLMcav-readout}). We now derive the simplified versions of these equations used in the main text to emphasise the essential physics of our scheme. 
We note that the cavity response matrix can be expressed in terms of the complex Lorentzian sideband amplitudes $\mathcal{L}(\Omega)\equiv (\kappa/2)/[\kappa/2 - i(\Omega+\Delta)]$ with phase $\Theta(\Omega)\equiv \text{Arg}[\mathcal{L}(\Omega)]$ as
\begin{align}
\mathbf{A}^{-1} &= \frac{1}{\kappa} \left(\begin{array}{cc}
    \mathcal{L}(\Omega) + \mathcal{L}^{\ast}(-\Omega) & i[\mathcal{L}(\Omega) - \mathcal{L}^{\ast}(-\Omega)] \\
    -i[\mathcal{L}(\Omega) - \mathcal{L}^{\ast}(-\Omega)] & \mathcal{L}(\Omega) + \mathcal{L}^{\ast}(-\Omega) 
\end{array}\right)\\
& =\frac{|\mathcal{L}(\Omega)|+|\mathcal{L}(-\Omega)|}{\kappa}e^{i[\Theta(\Omega)-\Theta(-\Omega)]/2}\mathbf{O}_{[\Theta(\Omega)+\Theta(-\Omega)]/2}\left[\mathbf{1}_{2}+i\frac{|\mathcal{L}(\Omega)|-|\mathcal{L}(-\Omega)|}{|\mathcal{L}(\Omega)|+|\mathcal{L}(-\Omega)|}\mathbf{O}_{-\pi/2}\right].\label{eq:Ainv-Polar}  
\end{align}
Assuming that the dependence of $\mathcal{L}(\Omega)$ on the Fourier frequency $\Omega$ is negligible over the bandwidth of interest, we may approximate $\mathcal{L}(\pm\Omega)\approx\mathcal{L}(\pm\omega_\text{M})$ (and accordingly $\Theta(\pm\Omega)\approx\Theta(\pm\omega_\text{M})$). Within this approximation, we can achieve the simplified mechanical response and input-output equations employed in the main text by introducing the rotated quadrature basis
\begin{equation}
    \bm{X}_{\text{L,M}}^{\text{in(ex)}\prime} \equiv e^{i[\Theta(\omega_\text{m})-\Theta(-\omega_\text{m})]/2}\mathbf{O}_{[\Theta(\omega_\text{m})+\Theta(-\omega_\text{m})]/2}\mathbf{O}_{\psi_\text{in}}^{\intercal}\bm{X}_{\text{L,M}}^{\text{in(ex)}}.\label{eq:X-LM-prime}
\end{equation}
In this way, using Eqs.~(\ref{eq:Ainv-Polar}) and (\ref{eq:X-LM-prime}) to reexpress the QBA force on the mechanical mode~(i.e., Eq.~(\ref{eq:Xm-chiEff}), 1st term in square brackets), we find
\begin{equation}
-\mathbf{C}\mathbf{A}^{-1}\mathbf{O}_{\psi_\text{in}}^{\intercal}\left(\sqrt{\kappa_{\text{in}}}\hat{{\bm{X}}}_{\text{L,M}}^{\text{in}}+\sqrt{\kappa_{\text{ex}}}\hat{{\bm{X}}}_{\text{L,M}}^{\text{ex}}\right) \approx 2\sqrt{\Gamma_\text{M}} \left(\begin{array}{c}
     1  \\
     i\zeta_\text{M} 
\end{array}\right)^{\intercal} \left(\sqrt{\kappa_{\text{in}}/\kappa}\hat{{\bm{X}}}_{\text{L,M}}^{\text{in}\prime}+\sqrt{\kappa_{\text{ex}}/\kappa}\hat{{\bm{X}}}_{\text{L,M}}^{\text{ex}\prime}\right),\label{eq:MQBA-simp}
\end{equation}
where we have introduced the mechanical readout rate and sideband asymmetry parameter,
\begin{align}
    \Gamma_\text{M} \equiv \frac{4g^2}{\kappa}(|\mathcal{L}(\omega_\text{M})|+|\mathcal{L}(-\omega_\text{M})|)^2,\quad
    \zeta_\text{M} \equiv \frac{|\mathcal{L}(\omega_\text{M})|-|\mathcal{L}(-\omega_\text{M})|}{|\mathcal{L}(\omega_\text{M})|+|\mathcal{L}(-\omega_\text{M})|},
\end{align}
respectively. Finally, we ignore the finite cavity overcoupling by setting $\kappa_\text{in}=\kappa$ (and hence $\kappa_\text{ex}=0$) in Eq.~(\ref{eq:MQBA-simp}) to arrive at the main-text expression for the response of $\hat{X}_\text{M}$. Noting that $-\mathbf{1}_2 + \kappa\mathbf{A}^{-1}=e^{i[\Theta(\omega_\text{m})-\Theta(-\omega_\text{m})]}\mathbf{O}_{\Theta(\omega_\text{m})+\Theta(-\omega_\text{m})}$, we find in the same limit that the rotated output quadrature
\begin{equation}
    \bm{X}_{\text{L,M}}^{\text{out}\prime}\equiv e^{-i[\Theta(\Omega)-\Theta(-\Omega)]/2}\mathbf{O}_{[\Theta(\Omega)+\Theta(-\Omega)]/2}^{\intercal}\mathbf{O}_{\psi_{\text{in}}}^{\intercal}\bm{X}_{\text{L,M}}^{\text{out}},
\end{equation}
obeys
\begin{equation}
    \bm{X}_{\text{L,M}}^{\text{out}\prime}=\bm{X}_{\text{L,M}}^{\text{in}\prime}+\sqrt{\Gamma_{\text{M}}}\left(\begin{array}{c}
i\zeta_{\text{M}}\\
1
\end{array}\right)\hat{X}_{\text{M}},\label{eq:IO-simp-appendix}
\end{equation}
as follows from $\hat{\bm{X}}_{\mathrm{L,M}}^{\textrm{out}}=-\hat{\bm{X}}_{\mathrm{L,M}}^{\textrm{in}}+\sqrt{{\kappa_\text{in}}}\hat{\bm{X}}_{\mathrm{L,M}}^\text{cav}$ combined with Eq.~(\ref{eq:XLMcav-readout}), again assuming $\kappa_\text{in}=\kappa$. Dropping the primes on the quadrature variables in Eq.~(\ref{eq:IO-simp-appendix}) for brevity we arrive at the input-output relation presented in the main text.

\subsection{Hybrid system I/O relations}
\label{app:model:hybrid}

The subsystems are coupled following the relation 
\begin{equation}
\hat{\bm{X}}_{\mathrm{L,M}}^{\textrm{in}}=\mathbf{O}_{\varphi} (\sqrt{\nu}\hat{\bm{X}}_{\mathrm{L,S}}^{\textrm{out}} + \sqrt{1-\nu}\hat{\bm{X}}_{\mathrm{L,}\nu}),\label{eq:XL-link}
\end{equation}
where optical transmission losses between the systems are modelled as a beam splitter with power transmission $\nu$, and $\hat{\bm{X}}_{\mathrm{L,S}}^{\textrm{out}}$ is defined Eq.~(\ref{eq:XLSfinal}). In general, the mechanical oscillator is not only coupled to light and its own thermal bath, but effectively also to the spin oscillator
\begin{multline}\label{eq:Xmfinal}
\hat{{X}}_{\mathrm{M}}^{\text{}}=-\chi_{\text{{M}00}}\mathbf{C}\mathbf{Y}^{-1}\mathbf{O}_{\psi_{\text{{in}}}}^{\intercal}\left(\sqrt{\nu\kappa_{\text{{in}}}}\mathbf{O}_{\varphi}[(\mathbf{1}_2 + 2\Gamma_{\text{S}}\mathbf{Z}\mathbf{L}\mathbf{Z})\hat{\bm{X}}_{\text{L,S}}^{\text{{in}}}+\sqrt{{\Gamma_{\text{S}}}}\mathbf{Z}\mathbf{L}\hat{{\bm{F}}}_{\text{S}}]
+\sqrt{(1-\nu)\kappa_{\text{{in}}}}\mathbf{O}_{\varphi} \hat{\bm{X}}_{\text{L},\nu}+\sqrt{{\kappa_{\text{ex}}}} \hat{\bm{X}}_{\text{{L,M}}}^{\text{{ex}}}\right)\\+(\chi_{\text{M}00}^{-1}-\mathbf{C}\mathbf{A}^{-1}\mathbf{B})^{-1}\hat{{F}}_{\text{M}},
\end{multline}
as follows by combining Eqs.~(\ref{eq:Xm-chiM00}), (\ref{eq:XL-link}), and (\ref{eq:XLSfinal}). 
At the output of the optical cavity, the field is homodyned at a quadrature of choice defined by the phase $\vartheta$, $\hat{{\bm{X}}}_{\mathrm{L}}^{\text{meas}} = \sqrt{\eta}\mathbf{O}_{\vartheta}\hat{\bm{X}}_{\mathrm{L,M}}^\textrm{out} +\sqrt{1-\eta}\hat{\bm{X}}_{\mathrm{L,}\eta}$, accounting for mode-matching and optical losses on the way to the final detector by the efficiency $\eta$. The detected field, including all contributions from losses, rotations and oscillator couplings is
\begin{multline}\label{eq:XLmeas}
\hat{{\bm{X}}}_{\mathrm{L}}^{\text{meas}}=\sqrt{{\eta}}\mathbf{O}_{\vartheta}\mathbf{\mathbf{O_{\psi_{\text{{out}}}}^{\intercal}}}(\kappa_{\text{{in}}}\mathbf{Y}^{-1}-\mathbf{1}_{2})\mathbf{O}_{\psi_{\text{{in}}}}^{\intercal}\left(\sqrt{\nu}\mathbf{O}_{\varphi}[(\mathbf{1}_{2} + 2\Gamma_{\text{S}}\mathbf{Z}\mathbf{L}\mathbf{Z})\hat{\bm{X}}_{\text{L,S}}^{\text{{in}}}+\sqrt{{\Gamma_{\text{S}}}}\mathbf{Z}\mathbf{L}\hat{{\bm{F}}}_{\text{S}}]+\sqrt{1-\nu}\mathbf{O}_{\varphi}\hat{\bm{X}}_{\text{L},\nu}\right)\\
+\sqrt{\eta\kappa_{\text{{in}}}{\kappa_{\text{ex}}}}\mathbf{O}_{\vartheta}\mathbf{O_{\psi_{\text{{out}}}}^{\intercal}}\mathbf{Y}^{-1}\mathbf{O}_{\psi_{\text{{in}}}}^{\intercal}\hat{\bm{X}}_{\text{{L,M}}}^{\text{{ex}}}-\sqrt{\eta\kappa_{\text{in}}}\mathbf{\mathbf{O}_{\vartheta}O_{\psi_{\text{{out}}}}^{\intercal}}\mathbf{Y}^{-1}\mathbf{B}\chi_{\text{{M}00}}\hat{{F}}_{\text{M}}+\sqrt{{1-\eta}}\hat{\bm{X}}_{\text{L,}\eta}.
\end{multline}
Note that the homodyne measurement only allows us to access one component of $\hat{{\bm{X}}}_{\mathrm{L}}^{\text{meas}}$ for a given choice of $\vartheta$.

The equations~(\ref{eq:XSfinal}), (\ref{eq:Xmfinal}), and (\ref{eq:XLmeas}) contain the full information needed to fit the experimental data and quantify correlations among the various constituents. To ease the handling of the theory, we construct a rectangular transformation matrix $\mathbf{U}$ in the input basis of the forces acting on the systems $\bm{Q}_\textrm{in} \equiv (\hat{F}_\text{S}^X, \hat{F}_\text{S}^P, \hat{F}_\text{M}, \hat{X}_\text{L,S}^\textrm{in}, \hat{P}_\text{L,S}^\textrm{in},\hat{X}_{\text{L},\nu}^\textrm{in}, \hat{P}_{\text{L},\nu}^\textrm{in},\hat{X}_\text{L,ex}^\textrm{in}, \hat{P}_\text{L,ex}^\textrm{in},\hat{X}_{\text{L},\eta}^\textrm{in}, \hat{P}_{\text{L},\eta}^\textrm{in})^\intercal$  such that 
\begin{align}
    \bm{Q}_\textrm{out}= \mathbf{U} \bm{Q}_\textrm{in}
\end{align}
and the output basis $\bm{Q}_\textrm{out}\equiv (\hat{X}_\text{M}, \hat{P}_\text{M}, \hat{X}_\text{S}, \hat{P}_\text{S}, \hat{P}_\text{L}^\textrm{meas})^\intercal$, which are all the output operators we might potentially be interested in.

The various power (and cross) spectral densities are calculated by taking the absolute square of the vector $\bm{Q}_\textrm{out}$ given the input matrix of spectral densities
\begin{equation}
\bar{\mathbf{S}}_\mathrm{in}\delta(\Omega-\Omega')=\frac{1}{2}\langle\bm{Q}^\dagger_\mathrm{in}(\Omega)[\bm{Q}_\mathrm{in}(\Omega')]^\intercal+\bm{Q}_\mathrm{in}(\Omega)[\bm{Q}_\mathrm{in}^\dagger(\Omega')]^\intercal\rangle,
\end{equation}
where $\intercal$ signifies a row-vector, while $\dagger$ indicates Hermitian conjugation of the individual vector elements, not the vector as a whole. $\bar{\mathbf{S}}_\mathrm{in}$ is a square matrix with diagonal entries
\begin{align}
    \textrm{diag}(\bar{\mathbf{S}}_\mathrm{in}) =\left(\overline{S}_{F_\textrm{S}^X F_\textrm{S}^X}, \overline{S}_{F_\textrm{S}^P F_\textrm{S}^P},\overline{S}_{F_\textrm{M} F_\textrm{M}},\overline{S}_{X_\textrm{L} X_\textrm{L}},\overline{S}_{P_\textrm{L} P_\textrm{L}},\overline{S}_{X_\textrm{L} X_\textrm{L}},\overline{S}_{P_\textrm{L} P_\textrm{L}}+\frac{\nu}{1-\nu}\overline{S}_\textrm{S,bb},\overline{S}_{X_\textrm{L} X_\textrm{L}},\overline{S}_{P_\textrm{L} P_\textrm{L}},\overline{S}_{X_\textrm{L} X_\textrm{L}},\overline{S}_{P_\textrm{L} P_\textrm{L}}\right), 
\end{align}
and all other elements equal to zero. Notably, for the easier theoretical treatment, the broadband noise is added via the inter-system loss port in the $\hat{P}_{\mathrm{L},\nu}^\mathrm{in}$ field. As defined above, it effectievly experiences same losses and rotation as the narrowband atomic noise. The various power spectral densities above are defined in \eqref{psdlight} and \eqref{psdthermalforces}, with Fourier frequency dependencies $\Omega$ dropped for clarity. The diagonal entries related to light variables are all vacuum, therefore the indistinguishable labelling.

This allows us to calculate the spectral densities of the output signals as follows
\begin{equation}
    \bar{\mathbf{S}}_\mathrm{out}=\mathbf{U}^\dagger \bar{\mathbf{S}}_\mathrm{in} \mathbf{U},\label{eq:S-bar-out}
\end{equation}
where $\mathbf{U}^\dagger$ is conjugate-transpose matrix w.r.t.\ to $\mathbf{U}$.
For instance, the (1,1) element of $\bar{\mathbf{S}}_\mathrm{out}$ is the power spectral density of the mechanical oscillator position $\bar{S}_{X_\mathrm{M}X_\mathrm{M}}$.

We may now calculate the steady-state unconditional covariance matrix in the spin-mechanics subspace. For this we integrate $\bar{\mathbf{S}}_\mathrm{MS}$, which we define as the submatrix of $\bar{\mathbf{S}}_\mathrm{out}$ containing the first 4 rows and columns, leading to the unconditional covariance matrix
\begin{equation}
    \mathbf{V}_u=\int_{-\infty}^\infty \frac{\mathrm{d}\Omega}{2\pi} \bar{\mathbf{S}}_\mathrm{MS}(\Omega).
\end{equation}
Figures \ref{fig:matrix}a, \ref{fig:matrix-epr}a and \ref{fig:matrix-det}a present examples of $\mathbf{V}_u$ in different cases and bases.

 \section{Wiener filtering} 
 \label{app:wiener}
 \begin{figure}
    \centering
    \includegraphics{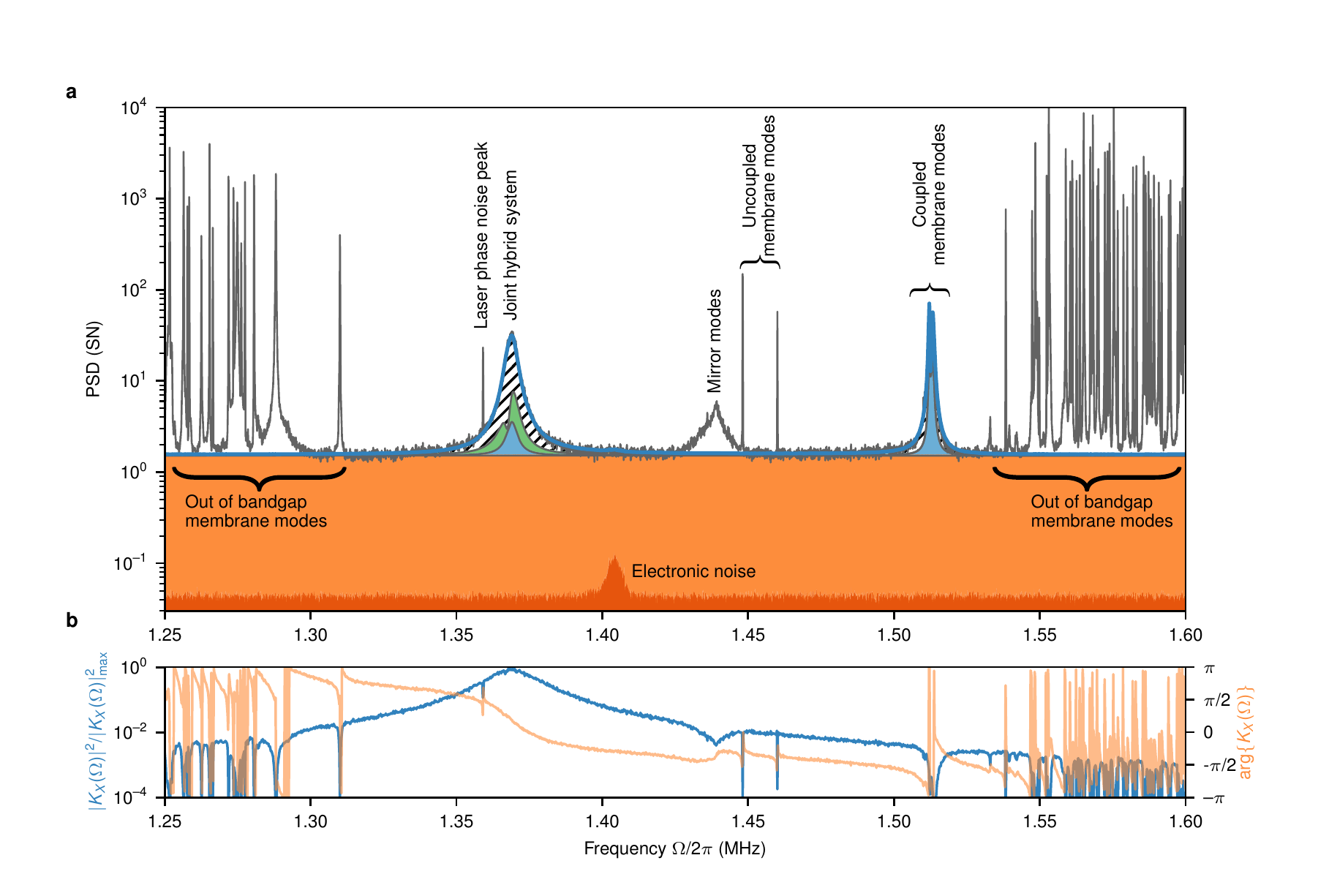}
    \caption{Spectra from Fig.~\ref{fig:spectra}b and associated Wiener filter from Fig.~\ref{fig:spectra}c shown in a wider range. \textbf{a,} the spectrum is again decomposed into the same components as in Fig.~\ref{fig:spectra}b, yet including the electronic detection noise (dark orange). Furthermore, additional membrane modes are visible in the experimental data. We model two of those modes, around 1.52 MHz, including their back-action. Those are the only high-$Q$ modes in the bandgap that are significantly coupled to light, which stretches from 1.31 to 1.54 MHz. Notably, the additional modes are treated as noise in the process of detecting the motion of the main defect mode of interest and the motion of spins. \textbf{b,} Wiener filter for the hybrid EPR system (squared normalized ampltide, left axis, phase, right axis). The filter automatically allows efficient tracking of the main EPR signal with other sources of noise removed in the form of frequency notch filters. Notably, the Wiener filter is significantly broader than the linewidth of the system itself. }
    \label{fig:WF-widerange}
\end{figure}

 A strong projective measurement of the initial system state $\hat{\rho}$ with a set of measurement operators $\{\hat{\Pi}_i\}$ generates a conditional quantum state $\hat{\rho}_c=\hat{\Pi}_i \hat{\rho} \hat{\Pi}_i/\Tr(\hat{\Pi}_i \hat{\rho} \hat{\Pi}_i)$. In a quasi-continuous (multi-step) weak measurement, we replace the projection operators with a set of generalized measurement operators (positive operator-valued measures) acting repeatedly on the initial state~\cite{Carmichael1993,Helstrom1976}. In general, prediction of the conditional state would require knowledge of operators associated with each measured value.
 For Gaussian states, the situation simplifies so that a linear, stationary filter can be used.
 
 Given the weak, continuous character of our optical probing, useful measurement results must necessarily be obtained as (weighted) averages over finite segments of the homodyne measurement current. 
 The appropriate temporal filter functions are defined by the system evolution during probing and the meter noise characteristics, necessitating precise knowledge of the equations of motion and the input-output relations. The methodology outlined here is known in classical physics and engineering as Kalman filtering, and its applicability to Gaussian quantum systems was proven in Refs.~\cite{YanbeiChenPRA2009,HaixingMiaoThesis} in a manner that we will now describe. 
 
 We note that our optical probing has the following two ``classical'' properties: First, the operators associated with the measurement current obtained at different times $t$ and $t'$ commute, $[\hat{P}_\textrm{L}^{\text{out}}(t),\hat{P}_\textrm{L}^{\text{out}}(t')]=0$, implying their \emph{simultaneous measurability}; second, causality entails that the measurement current at a given time $t$ does not respond to the future system evolution (at times $t'>t$), in turn leading to the property $[\hat{P}_\textrm{L}^{\text{out}}(t),\hat{X}(t')]=0$, $t'>t$, with $\hat{X}$ being any quadrature of a hybrid spin-mechanics system. Hence, the only manifestation of quantum mechanics in our probing scheme is that it enforces the presence of amplitude and phase (quantum) noise in the meter field according to the Heisenberg uncertainty relation. As the microscopic origin of the noise is immaterial to (classical) Wiener filtering theory, it follows from the above observations that it is applicable to our Gaussian quantum system.
 
 As necessary prerequisites we introduce the PSD of the measurement current as
 $\bar{S}_{ii}\equiv\bar{S}_{P_\mathrm{L}^\mathrm{meas}P_\mathrm{L}^\mathrm{meas}}$ since $i(t)$ is the result of a quadrature measurement. In general, we aim to track the entire hybrid system characterized by  $\bm{Q} = (\hat{X}_\textrm{M},\hat{P}_\textrm{M},\hat{X}_\textrm{S},\hat{P}_\textrm{S})^\intercal$. Furthermore, we need to consider the signal-current correlation (cross-spectral density) row vector
 $\bar{\mathbf{S}}_{\bm{Q}i}$, which is the last row of $\bar{\mathbf{S}}_\textrm{out}$, Eq.~(\ref{eq:S-bar-out}). 
The spectral densities $\bar{S}(\Omega)$ are used to compute temporal correlation functions $\bar{C}(\tau)$ using the inverse Fourier transform thanks to the Wiener-Khinchin theorem.
  
 Our hybrid system is driven solely by optical and thermal forces with wide-sense stationary noise statistics (i.e., constant first and second moments of all noises, and all covariances depending only on the time difference $t-t'$)~\citep{Broersen2006}. Under these circumstances the appropriate set of causal filters $\mathbf{K}$ for purposes of estimating the system first and second moments is the so-called \emph{Wiener filter} \cite{wiener1964extrapolation}. 
Convolving the filter with the measurement current yields the best unbiased estimate of the system variables (i.e., with the minimum mean-square error):
 \begin{equation}\label{eq:defQc}
     \bm{Q}^c_\infty(t) = \int_{-\infty}^t \mathbf{K}(t'-t) i(t')\, \mathrm{d}t',
 \end{equation}
 where $\bm{Q}^c = (X_\text{M}^c,P_\text{M}^c,X_\text{S}^c,P_\text{S}^c)^\intercal$ is the conditional trajectory in the steady-state scenario, i.e., for the case where we have $i(t)$ for all previous times available. In a more general case, where upon conditioning we increase the length of past data, we generally write:
  \begin{equation}
     \bm{Q}^c(t) = \int_{0}^t \mathbf{K}(t'-t,t) i(t')\, \mathrm{d}t',
     \label{eq:wpred-ft}
 \end{equation}
where $\mathbf{K}(\tau,t)$ is the filter function, $t'$ is the running argument of convolution and $t$ is the length of the conditioning interval.

To find the Wiener filters $\mathbf{K}$, we solve the Wiener-Hopf equations, which state that the optimal $\bm{Q}^{c}(t)$ must obey
\begin{equation}\label{eq:WH-gen}
\bar{\mathbf{C}}_{\bm{Q}^{c}i}(t')=\bar{\mathbf{C}}_{\bm{Q}i}(t'),    
\end{equation}
for all $t'$ within the conditioning window.
In the limit of infinite conditioning time, the Wiener-Hopf equation~(\ref{eq:WH-gen}) is typically stated as
\begin{equation}\label{eq:WH-infty}
      \int_{0}^\infty  \mathbf{K}^\intercal(-t'') \bar{C}_{ii}(t'-t'')\,\mathrm{d}t''=\bar{\mathbf{C}}_{\bm{Q}i}(t')\quad\forall\: t'\geq 0,
 \end{equation}
 where $\bar{\mathbf{C}}_{\bm{Q}i}(t')$ is the cross-correlation between $\bm{Q}$ and $i$ calculated as the inverse Fourier transform of $\bar{\mathbf{S}}_{\bm{Q}i}(\Omega)$, which is a row vector of cross-spectral densities (first four elements of the last row of $\bar{\mathbf{S}}_\mathrm{out}$, Eq.~(\ref{eq:S-bar-out})).
The vector form of the above equation should here be understood as 4 independent equations. 

If we only have data available for a finite past, we limit the above infinite integral to $t$ and find the finite-input response filter $\mathbf{K}(t',t)$ as a solution of
\begin{equation}\label{eq:defCQi}
      \int_{0}^t  \mathbf{K}^\intercal(-t'',t) \bar{C}_{ii}(t'-t'')\,\mathrm{d}t''=\bar{\mathbf{C}}_{\bm{Q}i}(t')\quad\forall\: t'\in[0,t].
 \end{equation}
 In this form, the Wiener-Hopf equation can also be easily discretised and cast in a matrix equation form. The solution is then obtained via the Levinson–Durbin recursion algorithm. 
 It is noteworthy that in the finite-time limit, the Wiener filter $\mathbf{K}(t',t)$ is only defined for $-t<t'<0$, in accordance with the integration domain in Eq.~(\ref{eq:wpred-ft}).

While the trajectory $\bm{Q}^c$ is stochastic, the variance of residual fluctuations is deterministic; it can be calculated as the difference between the \emph{unconditional} covariance matrix $\mathbf{V}_u$ and the (ensemble) covariance matrix of the best estimates $\mathbf{V}_\mathrm{be}$, 
 \begin{align}
    \label{eq:vcvuvbe}
     \mathbf{V}_c = \mathbf{V}_u - \mathbf{V}_\mathrm{be},
 \end{align}
 where
 \begin{align}\label{eq:vbe}
     \mathbf{V}_\mathrm{be} = \int_0^\infty \mathbf{K}(-t) \bar{\mathbf{C}}_{\bm{Q}i}(t)\, \mathrm{d}t=\text{Cov}(\bm{Q},\bm{Q}^{c}_\infty),
 \end{align}
is the 4 by 4 covariance matrix of the best estimates and $\bm{Q}^{c}(t)$ is given by Eq.~(\ref{eq:defQc}). 
In the case of a finite conditioning interval, we again limit the integration:
  \begin{equation}\label{eq:vbe-finite}
     \mathbf{V}_\mathrm{be}(t) = \int_0^t \mathbf{K}(-t',t) \bar{\mathbf{C}}_{\bm{Q}i}(t')\, \mathrm{d}t'=\text{Cov}(\bm{Q},\bm{Q}^{c}(t)),
 \end{equation}
 where $\bm{Q}^{c}(t)$ is defined by Eq.~(\ref{eq:wpred-ft}).
 Again, Eq.~(\ref{eq:vcvuvbe}) holds, and hence captures how the conditional variance evolves as we increase the conditioning time $t$.
 The relation $\mathbf{V}_\mathrm{be}(t)=\text{Cov}(\bm{Q},\bm{Q}^{c}(t))$ implied by Eq.~(\ref{eq:vbe-finite}) follows directly from the Wiener-Hopf equation~(\ref{eq:WH-gen}) by convolving it with $\mathbf{K}$ [as does the special case (\ref{eq:vbe})].
 
We present the obtained Wiener filter, for the point with best entanglement (see Methods~\ref{app:entanglement}), in Fig.~\ref{fig:spectra} of the main text, and for a wider frequency range in Fig.~\ref{fig:WF-widerange}. The time evolution of the conditional variance is shown in Fig.~\ref{fig:trajectory}, and the final $\Vc$ is shown in Fig.~\ref{fig:entanglement}.
 
Finally, it is noteworthy that the Wiener filter shares many characteristics with the widely used Kalman filter. In fact, the Wiener filter is a specific case of a Kalman filter where it can be obtained from the Wiener-Hopf equations since both noise and signal are wide-sense stationary. This still applies to our case of a finite-input response filter $\mathbf{K}(t',t)$, as to find it we assume stationary noise. In fact, the finite-input response (FIR) Wiener filters are widely used in engineering contexts. In a more general case one needs to solve Kalman equations that are qualitatively different.

\section{Entanglement estimation}
\label{app:entanglement} 
\begin{figure}[p]
    \centering
    \includegraphics[height=.28\textheight]{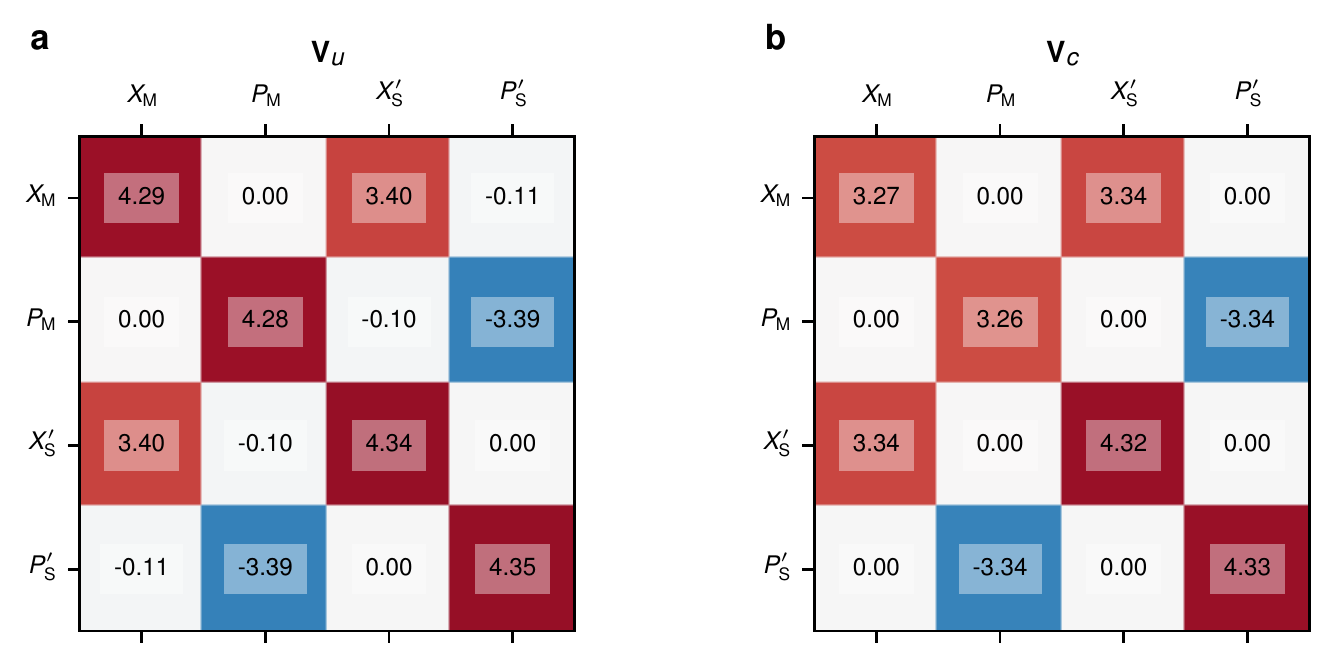}
    \caption{Covariance matrices in the individual single-system basis, for the dataset with $|\omega_\textrm{S}|-\omega_\textrm{M}\approx-\gamma_\textrm{M}/2$. Angle $\beta$ is adjusted so that anti-diagonal in \textbf{b} is 0.}
    \label{fig:matrix}

    \includegraphics[height=.28\textheight]{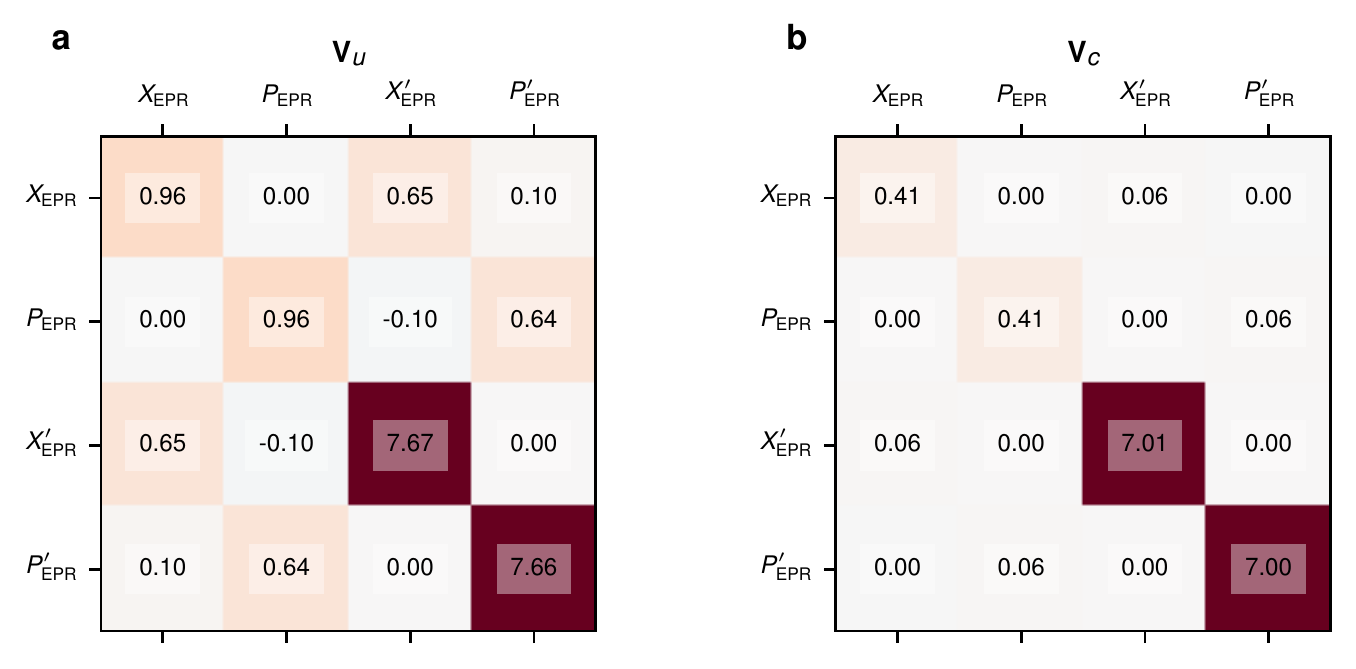}
    \caption{Covariance matrices in the EPR basis, for the dataset with $|\omega_\textrm{S}|-\omega_\textrm{M}\approx-\gamma_\textrm{M}/2$. \textbf{a} Unconditional and \textbf{b} conditional covariance matrices}
    \label{fig:matrix-epr}

    \includegraphics[height=.28\textheight]{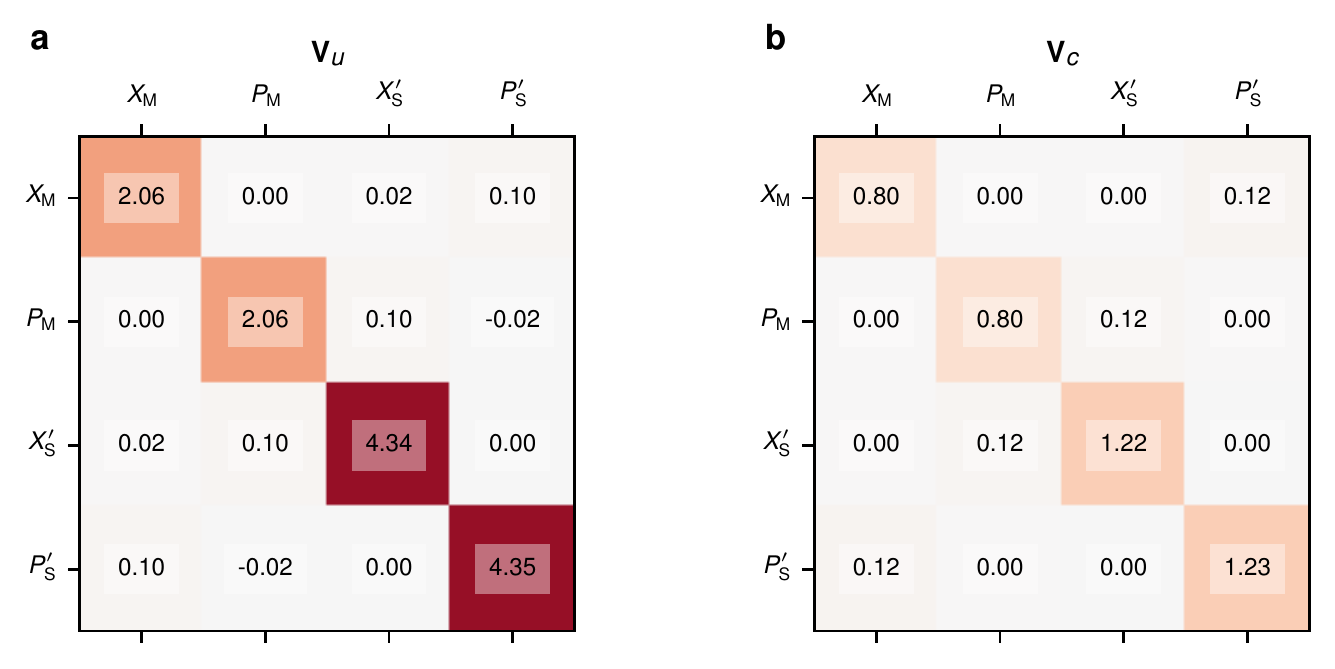}
    \caption{The covariance matrices in the detuned case, expressed in the individual system basis. Same rotation angle $\beta$ as in Fig \ref{fig:matrix} for the atomic subsystem is applied here}
    \label{fig:matrix-det}
\end{figure}

Let us now analyze the properties of the obtained covariance matrices and estimate the entanglement of the bipartite state. In Fig.~\ref{fig:matrix}(a) the covariance matrix $\mathbf{V}_u$ corresponding to the case with best entanglement is presented. Diagonal elements represent the occupations of individual oscillators. The conditioning procedure is then applied to obtain $\mathbf{V}_c$ in Fig.~\ref{fig:matrix}(b). Notably, we observe strong positive correlation between $\hat{X}_\text{M}$ and $\hat{X}'_\text{S}$ as well as negative correlation between $\hat{P}_\text{M}$ and $\hat{P}'_\text{S}$. Furthermore, we can see that the conditioning procedure mostly allow us to decrease the conditional occupation of the mechanical subsystem, which is most efficiently measured. The spin variables here are rotated, i.e., $\hat{X}'_\textrm{S}=\hat{X}_\mathrm{S} \cos\beta +\hat{P}_\mathrm{S} \sin\beta$, $\hat{P}'_\mathrm{S} =\hat{P}_\mathrm{S} \cos\beta -\hat{X}_\mathrm{S} \sin\beta$ such the anti-diagonal of $\mathbf{V}_c$ is nulled.

The estimation of the best entangled state involves the construction the general EPR variables
 \begin{gather}
     \hat{X}_\epr = (\hat{X}_\mathrm{M}- a \hat{X}'_\textrm{S})/\sqrt{1+a^2} = \bm{u}_X^\intercal \bm{Q},\\
     \hat{P}_\epr = (\hat{P}_\mathrm{M} + a \hat{P}'_\mathrm{S})/\sqrt{1+a^2} = \bm{u}_P^\intercal \bm{Q},\\
     \bm{\hat{X}}_\epr = \mathbf{u}^\intercal \bm{Q}
 \end{gather}
 (with matrix $\mathbf{u}$ having vectors $\bm{u}_X$ and $\bm{u}_P$ as columns) along with canonically conjugated variables
  \begin{gather}
     \hat{X}'_\epr = (\hat{X}_\mathrm{M}+ a \hat{X}'_\textrm{S})/\sqrt{1+a^2} \\
     \hat{P}'_\epr = (\hat{P}_\mathrm{M} - a \hat{P}'_\textrm{S})/\sqrt{1+a^2} 
 \end{gather}
 where $a$ is the relative weight of the spin component with respect to mechanics and $\beta$ is the rotation angle of the spin component, and $\bm{u}_X$ and $\bm{u}_P$ are unit-length vectors. The EPR variance (conditional or unconditional) $V=\Var[\hat{X}_{\epr}]+\Var[\hat{P}_{\epr}]$ is evaluated using the covariance matrix $\mathbf{V}$ as
 \begin{equation}
     V_{a,\beta} = \bm{u}_X^\intercal \mathbf{V} \bm{u}_X + \bm{u}_P^\intercal \mathbf{V} \bm{u}_P.
 \end{equation}
 For the present data, $a\approx 0.85$, which is approximately constant for all data point, and $\beta\approx\SI{20}{\degree}$ for the point of best entanglement. For different spin-mechanics detuning optimal $\beta$ varies by tens of degrees. We have minimized the EPR variance $V=\min_{a,\beta} V_{a,\beta}$ for both parameters individually for each dataset.
 
 Having defied the EPR basis we can now also plot the same matrix as in Fig.~\ref{fig:matrix} in the new basis, see Fig.~\ref{fig:matrix-epr}. Here, we observe that for $\mathbf{V}_c$ the variance of the EPR components on the diagonal indeed reaches below the classical limit of 0.5.
 
 Finally, we compared the entangled case with the far-detuned case, presented in Fig.~\ref{fig:matrix-det}. Here we observe negligible off-diagonal correlation terms, and also significantly lower unconditional occupation for mechanics, as it is not driven by the spin noise. Furthermore, the conditioning procedure can now distinguish the systems and efficiently brings down their respective conditional variances.
 
To generate the conditional trajectory in Fig.~\ref{fig:trajectory}(c) we first solve Eq.~(\ref{eq:defCQi}) for a set of conditioning times $t$ and find a collection of Wiener filters $\mathbf{K}(t',t)$. We then use the filters to get $\bm{Q}^c(t)$ as given by Eq.~(\ref{eq:wpred-ft}) as well as conditional covariance matrices $\mathbf{V}_c(t)$ (see Eqs.~(\ref{eq:vcvuvbe}) and (\ref{eq:vbe-finite})). We then find the optimal $a$ and $\beta$ for the $\mathbf{V}_c$ associated with $t\rightarrow\infty$, which gives us $\mathbf{u}$. Subsequently, $\bm{X}_\epr^c=\mathbf{u}^{\intercal}\bm{Q}^c$ is calculated. Finally we move to a rotating frame by $\bm{\tilde{X}}_\epr^c=\mathbf{O}_{\omega t} \bm{X}_\epr^c$ with $\omega/2\pi=\SI{1.37}{\MHz}$, which is rather an arbitrary choice since for the EPR oscillator there is no single distinguished frequency unless $\omega_\textrm{M}=|\omega_\text{S}|$ exactly, which is not the case.

We observe $\Var[\hat{X}]\approx\Var[\hat{P}]$ for all cases (Figs.~\ref{fig:matrix}--\ref{fig:matrix-det}) consistent with our system operating within the regime of validity for the Rotating Wave Approximation.

\section{Uncertainties}
\label{app:mcmc}
We apply elaborate statistical techniques to deduce the statistical uncertainty for the value of the degree of entanglement.

Spectra corresponding to points in Fig.~\ref{fig:entanglement}b are fitted collectively to the same model. A subset of the parameters is shared between all spectra, while others are allowed to fluctuate from spectrum to spectrum, representing small short-timescale fluctuations. 
 
 The parameters that are allowed to change from spectrum to spectrum are atomic frequency, \LO1+\LO2 phase $\varphi$, cavity detuning $\Delta$, and mechanical coupling rate $g$. The drift of the latter two can be explained by a spurious interference, which turns drifts in $\varphi$ into drifts in optical power in \LO2, which in turns leads to a change in $\Delta$ and thus also $g$. The typical size of drifts of $\varphi$ is $\sim3$ degrees.
 
We establish prior probabilities for all parameters by independent measurements and calibrations, many of which we explain above. We use those priors for our parameters, together with the spectra and their statistical uncertainties to perform a log-likelihood optimisation. We use Gaussian priors for the parameters and assume a relative Gaussian error of \SI{8}{\percent}, stemming from the number of samples for each spectrum $N_\mathrm{samp}=200$, i.e., the statistical variance of the periodogram estimator, and additional uncertainty due to shot-noise level calibration. We additionally assume the level of data uncertainty to have an extra constant offset of 0.1 SN units to account for the presence of small mirror mode peaks beneath the signal. 

Due to the vast parameter space, originating partly from the collective fitting with both shared and non-shared parameters, we perform the optimisation with Markov Chain Monte Carlo (MCMC) simulations \cite{emcee}. We run 150 walkers with 4000 burn-in steps and subsequent 6000 sampling steps. From these \num{900000} points, we select 1000 random samples for which we compute entanglement. This sampling of the log-likelihood landscape leads directly to posterior probabilities for the parameters for each spectrum, and, more importantly, also for derived values, such as the conditional variance. The choice of the number of samples for the entanglement calculation is determined by the computational cost of evaluating the conditional variance. Sampling those \num{1000} points from a larger set of MCMC points reduces the co-variance of the points sampling of the posterior log-likelihood landscape.

The MCMC fitting routine results in a set of parameters with good agreement between priors and posteriors for almost all parameters. The main discrepancy is for the case of inter-system quantum efficiency; here, the posterior value of $\nu = \effnu$ is significantly lower than the anticipated value of $\nu_\mathrm{prior} = 0.65\pm0.03$. In addition, we obtain a slightly lower posterior detection efficiency $\eta=\effeta$ than $\eta_\mathrm{prior}=0.80\pm0.03$ and higher overcoupling $(\kappa_\mathrm{in}/\kappa)=\num{0.925\pm.005}$ 
than $(\kappa_\mathrm{in}/\kappa)_\mathrm{prior}=0.91\pm0.01$. The extra optical losses are currently unaccounted for, with possible explanations for this discrepancy that include mode matching and polarisation-dependent losses of our quantum signal. We should stress that this discrepancy leads only to a reduction of the obtained entanglement. The atomic parameters are kept reasonably within the prior bounds with $\Gamma_\mathrm{S,prior}=\SI{18\pm1}{\kilo\hertz}$ and posterior $\Gamma_\mathrm{S}=\SI{20.3\pm0.4}{\kilo\hertz}$ as well as $n_\mathrm{S,prior}=0.72\pm0.05$ and posterior $n_\mathrm{S}=0.81\pm0.05$.

\clearpage
\onecolumngrid

\begin{table}
\begin{tabular}{SSS} 
    {\textbf{Parameter}} & {\textbf{\text{Symbol}}} & {\textbf{Value}} \\ \midrule
     & {\textbf{Atomic spin oscillator}} &  \\
    {Decoherence rate in the dark}  & {$\gamma_{\text{S0,dark}}/2\pi$} & {\SI{450}{Hz}}  \\
        {Intrinsic linewidth}  & {$\gamma_{\text{S0}}/2\pi$} & {\gammaSzero}  \\
    {Effective linewidth (incl.\ dynamical damping)}  & {$\gamma_{\text{S}}/2\pi$} & {\gammaS} \\
    {Tensor contribution}  & {$\zeta_\mathrm{S}$} & {\zetaS} \\
    {LO$_1$ driving power}  &   &  {\SI{350}{\micro\watt}} \\
    {Readout rate}  & {$\Gamma_{\textrm{S}}/2\pi$} &  {$\GammaS$} \\
    {Spin Polarization}    & {$p$}  & {0.82} \\
    {Spin thermal occupancy}  & {$n_\text{S}$} &{\nS}\\
    {Microcell temperature}    &  & {50$^\circ\text{C}$}  \\
    \midrule
    
    & {\textbf{Mechanical oscillator and cavity}} & \\
    {Intrinsic mechanical frequency}  & {$\omega_{\text{M0}}/2\pi$} & {\SI{\Omegamzero}{MHz}} \\
    {Intrinsic damping rate}  & {$\gamma_{\text{M0}}/2\pi$} & {\SI{\gammamechnat}{mHz}} \\
    {Optical damping rate} & {$\gamma_{\text{M}}/2\pi$}  & {$\gammaM$}\\
    {Cavity detuning}  & {$\Delta/2\pi$}  & {\avgDelta}  \\
    {Total cavity linewidth}  & {$\kappa/2\pi$} & {\SI{\kappamech}{\mega\hertz} }\\
    {{\LO2} drive power}  & { } & \SI{\sim8}{\micro\watt}\\ 
    {Intracavity photons} & $N$ & {1.6$\times$10$^\text{6}$}\\
    {Single photon coupling rate}  & {$g_0/2\pi$} & {\SI{6e1}{\hertz}}\\
    {Readout rate}  & {$\Gamma_{\textrm{M}}/2\pi$} &  {$\GammaM$} \\
    {Cavity overcoupling} & {$\kappa_\mathrm{in}/\kappa$} & {0.\ovc} \\
    {Thermal bath temperature} & {$T$} & $\SI{\mechT}{\kelvin}$ \\
    {Bath occupancy}  & {$n_\text{M0}$}  & {$\nthmech\times$10$^\text{3}$} \\
    {Mean occupancy}  & {$n_\text{M}$} &{\nM} \\
    {Quantum cooperativity} & {$\Cq^\text{M}$}  &  {$\CqM$} \\

    \midrule
    
    & {\textbf{Hybrid \& detection}} &  \\
    {Quantum efficiency between systems}  & {$\nu$} & {$\effnu$} \\
    {Cavity mode-matching (amplitude)} & {} & {0.9}\\
    {Power transmission between systems} & {} & {0.8} \\
    {Detection efficiency}  & {$\eta$} & {$\effeta$} \\
    {Homodyning visibility}  & { }  & {0.96} \\
    {Power transmission and detector QE} & {} & {0.87} \\
    {\LO1--\LO2 phase} & {$\varphi$} & {\SI{\sim180}{\degree}}\\
    {Detection phase} & {$\vartheta$} & {\SI{2}{\degree}}\\
    \bottomrule
    
\end{tabular}
\caption{ \textbf{Summary of notation and experimental parameters.} When applicable, we quote the posterior mean values from the MCMC simulation.}
\label{table:parameters}
\end{table}

\end{document}